\documentclass[aps,preprint,showpacs]{revtex4}
\usepackage{graphicx,amsbsy}
\begin{document}
\title{Robust generation of entanglement in Bose-Einstein condensates by 
collective atomic recoil} 
\author{Mary M. Cola, Matteo G. A. Paris and Nicola Piovella}
\affiliation{Dipartimento di Fisica dell'Universit\`a di Milano}
\affiliation{I.N.F.N. \& I.N.F.M. @ Universit\`a di Milano, Via Celoria 16, 
Milano I-20133,Italy}
\begin{abstract}
We address the dynamics induced by collective atomic recoil in a Bose-Einstein
condensate in presence of radiation losses and atomic decoherence. In
particular, we focus on the linear regime of the lasing mechanism, and analyze
the effects of losses and decoherence on the generation of entanglement. The
dynamics is that of three bosons, two atomic modes interacting with a
single-mode radiation field, coupled with a bath of oscillators. The resulting
three-mode dissipative Master equation is solved analytically in terms of the
Wigner function.  We examine in details the two complementary limits of {\em
high-Q cavity} and {\em bad-cavity}, the latter corresponding to the so-called
superradiant regime, both in the quasi-classical and quantum regimes.  We
found that three-mode entanglement as well as two-mode atom-atom and
atom-radiation entanglement is generally robust against losses and
decoherence,thus making the present system a good candidate for the
experimental observation of entanglement in condensate systems. In particular,
steady-state entanglement may be obtained both between atoms with opposite momenta 
and between atoms and photons.
\end{abstract} 
 \maketitle
\section{Introduction}
The experimental realization of Bose-Einstein condensation opened the
possibility to generate macroscopic atomic fields whose quantum statistical
properties can in principle be manipulated and controlled \cite{MeyBOOK}. The
system considered here to this purpose is an elongated Bose-Einstein
Condensate (BEC) driven by a far off-resonant pump laser of  wave vector
$k_p=\omega_p/c$ along the condensate long axis and coupled to a single mode in an
optical ring cavity. The mechanism at the basis of this kind of physics is the
so-called Collective Atomic Recoil Lasing (CARL)\cite{CARL} in his full
quantized version \cite{Moore:1,Moore:2,PRA}. In CARL the scattered radiation
mode and the atomic momentum side modes become macroscopically occupied via a
collective instability. A peculiar aspect of the quantum regime is the
possibility of populating single momentum modes separated by $\Delta p=2\hbar
k_p$ off the condensate ground state with zero initial momentum.
The experimental observation of CARL in a BEC has been until now realized in
the so-called superradiant regime \cite{MIT,Tokio,LENS}, {\em i.e.} 
without the optical cavity. In this case the radiation is emitted along the 
'end-fire modes' of the condensate \cite{Moore:3} with very large radiation 
losses (in the mean field model, with $\kappa\approx c/L$, where $\kappa$ is 
the cavity decay rate and $L$ is the condensate length). In a recent work \cite{PRA} 
it has been shown that atom-atom and atom-photon entanglement can be produced 
in the linear regime of CARL, in which the ground state of the condensate remains 
approximately undepleted. In this regime the atomic multi-mode system can be 
described by only two momentum side modes, with $p=\pm 2\hbar k_p$. This source of
entanglement has been also proposed for a quantum teleportation scheme among
atoms and photons \cite{telebec}. The results presented in \cite{PRA} refer to 
the ideal case of a perfect optical cavity and an atomic system free of decoherence. 
However, in view of an experimental observation of entanglement, a detailed
analysis of the sources of noise is in order, which in turn may be a serious 
limitation for entanglement in CARL \cite{Gasenzer}. Also, it has not yet been
proved that BEC superradiance experiments may generate entangled atom-photon
states, as suggested in \cite{Moore:3}. This issue is investigated for the
first time in this paper, where we demonstrate the entangled properties of the
atom-atom and atom-photon pairs produced in the linear stage of the
superradiant CARL regime in a BEC.
\par
The aim of the present work is to analyze systematically, by solving the 
three-mode Master equation in the Wigner representation, the effects 
of losses and decoherence on the generation of entanglement.
We will first investigate the effects of either a small atomic decoherence or a
finite mirror transmission of the optical cavity, and then analyze in details 
the generation of entanglement in the superradiant regime, where the cavity 
losses are important. 
\par
The paper is structured as follows. In Section \ref{s:me} we briefly review
the ideal dynamics and derive the general solution of the Master equation. 
In Section \ref{s:vac} we consider the evolution of the system starting from 
the vacuum and calculate the relevant expectation values, such as average and 
variance of the occupation number and two-mode squeezing parameters.
In Section \ref{s:wr} the different working regimes are introduced and
the dynamics analyzed, whereas in Section \ref{s:en} we investigate three- 
and two-mode entanglement properties of the system as a function of loss and 
decoherence parameters. Section \ref{s:out} closes the paper with some
concluding remarks. 

\section{Dissipative Master Equation}\label{s:me}

We consider a $1$D geometry in which a off-resonant laser pulse, with Rabi
frequency $\Omega_0=dE_0/\hbar$ (where $d$ is the dipole matrix element and
$E_0$ is the electric field amplitude) and detuned from the atomic resonance
by $\Delta_0=\omega_p-\omega_0$, is injected in a ring cavity aligned with 
the symmetry $z$-axis of an elongated BEC.  
The dimensionless position and momentum of the atom along the
axis $\hat z$ are $\theta=2k_pz$ and $p=p_{z}/2\hbar k_p$.  The interaction time
is $\tau=\rho\omega_r t$, where $\omega_r=2\hbar k_p^2/m$ is the recoil
frequency, $m$ is the atomic mass,
$\rho=\left(\Omega_0/2\Delta_0\right)^{2/3}\left(\omega_p d^2
N/V\hbar\epsilon_0\omega_r^2\right)^{1/3}$ is the CARL parameter, $N$ is the
number of atoms in the cavity mode volume $V$ and $\epsilon_0$ is the
permittivity of the free space.
\par
In a second quantized model for CARL \cite{PRA,Moore:2} the atomic field
operator $\hat\Psi(\theta)$ obeys the bosonic equal-time commutation relations
$[\hat\Psi(\theta),\hat\Psi^{\dag}(\theta')]=\delta(\theta-\theta')$,
$[\hat\Psi(\theta),\hat\Psi(\theta')]=0$ and the normalization condition is
$\int_0^{2\pi}d\theta\hat\Psi(\theta)^{\dag}\hat\Psi(\theta)=N$.  We assume
that the atoms are delocalized inside the condensate and that, at zero
temperature, the momentum uncertainty $\sigma_{p_z}\approx\hbar/\sigma_z$ can
be neglected with respect to $2\hbar k_p$. This approximation is valid for
$L\gg\lambda_p$, where $L$ is the condensate length and $\lambda_p=2\pi/k_p$ is the laser
radiation wavelength. In this limit, we can introduce creation and
annihilation operators for an atom with a definite momentum $p$, i.e.
$\hat\Psi(\theta)=\sum_m \hat c_m\langle\theta|m\rangle$, where
$p|m\rangle=m|m\rangle$ (with $m=-\infty,\dots,\infty$),
$\langle\theta|m\rangle=(1/\sqrt{2\pi})\exp(im\theta)$ and $\hat c_m$ are bosonic
operators obeying the commutation relations $[\hat c_m,\hat c^{\dag}_n]=\delta_{mn}$
and $[\hat c_m,\hat c_n]=0$.  The Hamiltonian in this case is \cite{PRA}
\begin{equation}
\hat H= \sum_{n=-\infty}^\infty\left\{
\frac{n^2}{\rho}\hat c_n^{\dag}\hat c_n+i\sqrt{\frac{\rho}{2N}}\left(\hat a^{\dag}
\hat c_{n}^{\dag}\hat c_{n+1}-{\rm
h.c.}\right)\right\} -\delta \hat a^{\dag}\hat a \label{ham2}
\end{equation}
where $\hat a$ is the annihilation operator (with $[\hat a,\hat a^{\dag}]=1$)
for the cavity mode (propagating along the positive direction of the $z$-axis)
with frequency $\omega_s$ and $\delta=(\omega_p-\omega_s)/\rho\omega_r$ is the
detuning with respect to the pump frequency $\omega_p$.  Let us now consider the
equilibrium state with no photons and all the atoms at rest, i.e. with
$|\Psi_0\rangle=\sqrt{N}|0\rangle$. Linearizing around this equilibrium state
and defining the operators $\hat a_1=\hat c_{-1}e^{i\delta\tau}$, $\hat
a_2=\hat c_1e^{-i\delta\tau}$ and $\hat a_3=\hat a e^{-i\delta\tau}$, the
Hamiltonian (\ref{ham2}) reduces to that for three parametrically coupled
harmonic oscillator operators:
\begin{equation}
\hat{H}=\delta_+ \hat{a}_2^{\dag}\hat{a}_2 -\delta_-
\hat{a}_1^{\dag}\hat{a}_1 +i\sqrt{\frac{\varrho}{2}}
\;\left[(\hat{a}_1^{\dag}+\hat{a}_2)\hat{a}_3^{\dag}-
(\hat{a}_1+\hat{a}_2^{\dag})\hat{a}_3\right],\label{ham}
\end{equation}
where $\delta_{\pm}=\delta\pm 1/\rho$.
In Ref.\cite{PRA} we have explicitly evaluated the state evolved from
the vacuum of the three modes, $|0_1,0_2,0_3\rangle$, as
\begin{eqnarray}
|\psi(\tau)\rangle=\frac{1}{\sqrt{1+\langle \hat{n}_1\rangle}}
\sum_{n,m=0}^\infty
\left(\frac{ \langle \hat{n}_3\rangle}{1+
\langle \hat{n}_1\rangle}\right)^{m/2}
\left(\frac{ \langle \hat{n}_2\rangle}{1+
\langle \hat{n}_1\rangle}\right)^{n/2}
e^{-i(n\phi_2+m\phi_3)}
\sqrt{\frac{(m+n)!}{m!n!}}
|m+n,n,m\rangle
\label{Tstate}\;,
\end{eqnarray}
where $\langle \hat n_i\rangle=\langle \hat a^\dag_i\hat a_i\rangle$ with $i=1,2,3$ are the 
expectation values of the occupation numbers of the three modes, related by the constant of motion
$\hat C=\hat n_1-\hat n_2-\hat n_3$.
In this paper we extend our previous analysis to include the effects of
atomic decoherence and cavity radiation losses. In this case the dynamics
of the system in described by the following Master equation:
\begin{eqnarray}
\frac{d\hat{\varrho}}{d\tau}= -i
\left[\hat{H},\hat{\varrho}\right]+2\gamma_1
L[\hat{a}_{1}]\hat{\varrho}+2\gamma_2 L[\hat{a}_{2}]\hat{\varrho}+2\kappa
L[\hat{a}_{3}]\hat{\varrho} \label{ME},
\end{eqnarray}
where $\gamma_1$, $\gamma_2$ and $\kappa$ are the damping rates for the modes $a_i$ and
$L[\hat{a}_i]$ is the Lindblad superoperator
\begin{equation}
L[\hat{a}_i]\hat{\varrho}=\hat{a}_i\hat{\varrho}
\hat{a}_i^{\dag}-\frac12 \hat{a}_i^{\dag}\hat{a}_i\hat{\varrho} -
\frac12 \hat{\varrho} \hat{a}_i^{\dag} \hat{a}_i.
\end{equation}
The atomic decay stems from coherence loss between the undepleted ground
state with $p_z=0$ and the side modes with $p_z=\pm 2\hbar k_p$. In general, we
assume that the two atomic modes may have different decoherence rates,
depending on the direction of recoil \cite{LENS}. The radiation decay constant
is $\kappa=cT/2\cal{L}$, where $T$ is the transmission of the cavity and $\cal
{L}$ is the cavity length.  Through a standard procedure \cite{Carmichael},
the Master equation can be transformed into a Fokker-Planck equation for the
Wigner function of the state $\hat{\varrho}$,
\begin{equation}
W(\alpha_{1},\alpha_{2},\alpha_{3},\tau)=\int\prod_{i=1}^{3}
\frac{d^{2}\xi_i}{\pi^2}
\;e^{\xi_{i}^{*}\alpha_{i}-\alpha_{i}^{*}\xi_{i}}
\chi(\xi_1,\xi_2,\xi_3,\tau)\:, \label{wignerdef}
\end{equation}
where $\alpha_j$ and $\xi_j$ are complex numbers 
and $\chi$ is the characteristic function defined as
\begin{eqnarray}
\chi(\xi_1,\xi_2,\xi_3) &=& \hbox{Tr}\Big[\hat{\varrho}\:
\hat{D}_1(\xi_1) \hat{D}_2(\xi_2) \hat{D}_3(\xi_3) \Big]
\label{chfun},
\end{eqnarray}
where $\hat{D}_j(\xi_j)=\exp(\xi_j\hat{a}^\dag_j - \xi^{*}_j \hat{a}_j)$ is a
displacement operator for the $j$-th mode.  Using the differential
representation of the Lindblad superoperator, the Fokker-Planck equation is:
\begin{eqnarray}
\frac{\partial W}{\partial \tau}=
-\left(\textbf{u}'^{T}\mathbf{A}\textbf{u} +{\rm c.c.}\right)W
+\textbf{u}'^{T}\mathbf{D}\textbf{u}'^{*}W\label{fokker}
\end{eqnarray}
where
\begin{equation}
\textbf{u}^T= \left( \alpha_{1}^{*}, \alpha_{2}, \alpha_{3}
\right) \qquad \textbf{u}'^{T}=\left(
\frac{\partial}{\partial\alpha_{1}^{*}},
 \frac{\partial}{\partial \alpha_{2}},
\frac{\partial}{\partial \alpha_{3}}\right)
\end{equation}
and $\mathbf{A}$ and $\mathbf{D}$ are  the following
drift and diffusion matrices:
\begin{equation}
\mathbf{A}=\left(\begin{array}{ccc}
\gamma_1+i\delta_{-} & 0 & -\sqrt{\rho/2}  \\
0 &\gamma_2+i\delta_{+} & \sqrt{\rho/2}  \\
-\sqrt{\rho/2} & -\sqrt{\rho/2} & \kappa
\end{array}\right)
\qquad\mathbf{D}=\left(\begin{array}{ccc}
 \gamma_1& 0& 0\\
 0&\gamma_2&0 \\
 0&0 &\kappa
\end{array}\right)\:.
\end{equation}
\par
The solution of the Fokker-Planck equation (\ref{fokker}) reads as follows 
\begin{equation} 
W(\mathbf{u},\tau)=\int d^{2}\mathbf{u}_{0}\:
W(\mathbf{u}_{0},0)\:G(\mathbf{u},\tau;\mathbf{u}_{0},0)\label{conv} \end{equation} where
$W(\mathbf{u}_{0},0)$ is the Wigner function for the initial state and the
Green function $G(\mathbf{u},t;\mathbf{u}_{0},0)$ is the solution of
Eq.(\ref{fokker}) for the initial condition
$G(\mathbf{u},0;\mathbf{u}_{0},0)=\delta^{(3)}(\mathbf{u}-\mathbf{u}_{0})$.
The calculation of the Green function, solution of Eq.(\ref{fokker}), is
reported in detail in Appendix A and yields the following result:
\begin{equation}
G(\mathbf{u},\tau;\mathbf{u}_{0},0)=\frac{1}{\pi^{3}\det\mathbf{Q}(\tau)}
\exp\left\{-\left[\mathbf{u}-\mathbf{M}(\tau)\mathbf{u}_{0}\right]^\dag
\mathbf{Q}^{-1}(\tau)\left[\mathbf{u}-\mathbf{M}(\tau)\mathbf{u}_{0}\right]\right\}.
\label{Green}
\end{equation}
where
\begin{equation}
\mathbf{M}(\tau)\equiv e^{\mathbf{A}\tau}=
\left(\begin{array}{rrc}
 f_{11}(\tau) & f_{12}(\tau)& f_{13}(\tau)\\
-f_{12}(\tau) & f_{22}(\tau)& f_{23}(\tau)\\
 f_{13}(\tau) &-f_{23}(\tau)& f_{33}(\tau)
\end{array}\right)
\label{eAt}
\end{equation}
and
\begin{equation}
\mathbf{Q}(\tau)=\int_0^\tau d\tau' \: \mathbf{M}(\tau')\:\mathbf{D}\:
\mathbf{M}^\dag(\tau')\:.\label{Q}
\end{equation}
In Eq.(\ref{eAt}) the complex functions $f_{ij}$, given explicitly in Appendix B,
are the sum of three terms proportional to $e^{i\omega_k\tau}$, where $\omega_k$,
with $k=1,2,3$, are the three roots of the cubic equation:
\begin{equation}
\left[\omega-\delta-i\left(\kappa-\gamma_+\right)\right]
\left[\omega^2-\left(\frac{1}{\rho}+i\gamma_-\right)^2\right]
+1+i\rho\gamma_-=0
\label{cubica}
\end{equation}
and $\gamma_\pm=(\gamma_1\pm\gamma_2)/2$.

\section{Evolution from vacuum and expectation values}\label{s:vac}

Let now assume that the initial state is the vacuum. The  
characteristic function and the Wigner function at $\tau=0$ are given by
\begin{eqnarray}
\chi({\boldsymbol \xi}) = \exp \left\{-{\boldsymbol \xi}^\dag  {\mathbf C}_0{\boldsymbol \xi}
\right\} \qquad
W(\mathbf{u},0) = \left(\frac{2}{\pi}\right)^3 \exp\left\{-
\mathbf{u}^\dag{\mathbf C}_0^{-1} \mathbf{u}\right\}.
\label{W0}\;,
\end{eqnarray}
where ${\boldsymbol \xi}=(\xi_1^*,\xi_2,\xi_3)$ and the covariance matrix is
multiple of the identity matrix ${\mathbf C}_0=\frac12 {\mathbf I}$.
Since the initial state is Gaussian and the convolution in (\ref{conv})
maintains this character we have that the Wigner function is Gaussian at any time 
$\tau$. After some algebra, we found that the covariance matrix is given by
\begin{equation}
\mathbf{C}(\tau)=\mathbf{Q}(\tau)+\frac{1}{2}\mathbf{M}(\tau)\mathbf{M}^{\dag}(\tau),
\label{cov}
\end{equation}
where the explicit form of the elements $C_{ij}=\langle (u_i-\langle u\rangle_i)(u_j-\langle u
\rangle_j)^*\rangle$ in terms of the functions $f_{ij}$ is reported in appendix
\ref{a:appB}. Since the state is Gaussian, from (\ref{cov}) it is possible to derive
all the expectation values for the three modes. In particular, $C_{ii}=1/2+\langle{\hat
n_i}\rangle$, 
$C_{12}=\langle{\hat a^\dag}_1 {\hat a^\dag}_2\rangle$, 
$C_{13}=\langle{\hat a^\dag}_1{\hat a^\dag}_3\rangle$ and
$C_{23}=\langle{\hat a}_2{\hat a^\dag}_3 \rangle$.
The number variances and the equal-time correlation functions for the mode
numbers are calculated from the forth-order covariance matrix
$G_{ijkl}=\langle(u_i-\langle u\rangle_i)(u_j-\langle u\rangle_j)
(u_k-\langle u\rangle_k)^*(u_l-\langle u\rangle_l)^*\rangle$, which in turn
is related to covariance matrix as follows: 
\begin{equation}
G_{ijkl}=C_{ki}C_{lj}+C_{li}C_{kj}\:.
\label{G}
\end{equation}
In particular, we have
\begin{eqnarray}
G_{iiii}&=&\langle{\hat n_i^2}\rangle+\langle{\hat n_i}\rangle+\frac{1}{2}\label{G11}\\
G_{ijij}&=&\langle{\hat n_i\hat n_j}\rangle+\frac{1}{2}\langle{\hat n_i}\rangle
+\frac{1}{2}\langle{\hat n_j}\rangle+\frac{1}{4}\qquad (i\neq j)\label{G12}
\end{eqnarray}
From Eqs.(\ref{G})-(\ref{G12}) it follows that:
\begin{eqnarray}
\sigma^2(n_i)&=&
\langle{\hat n_i}\rangle(\langle{\hat n_i}\rangle+1)\label{var}\\
g_{i}^{(2)}&=&\frac{\langle\hat a_i^{\dag}\hat a_i^{\dag}\hat a_i\hat a_i\rangle}
{\langle\hat n_i\rangle^2}=2\label{corr1}\\
g_{i,j}^{(2)}&=&\frac{\langle\hat n_i\hat n_j\rangle}
{\langle\hat n_i\rangle\langle\hat n_j\rangle}=1+\frac{|C_{ij}|^2}
{\langle\hat n_i\rangle\langle\hat n_j\rangle},\label{corr2}
\end{eqnarray}
where $\sigma^2(n_i)=\langle\hat n_i^2\rangle-\langle{\hat n_i}\rangle^2$,
with $i=1,2,3$, and $i\neq j$ in Eq.(\ref{corr2}). 
The two-mode number squeezing parameter is calculated as \cite{Burnett}:
\begin{equation}
\xi_{i,j}=\frac{\sigma^2(\hat n_i-\hat n_j)}{\langle\hat n_i\rangle +\langle\hat n_j\rangle}
=\frac{\sigma^2(n_i)+\sigma^2(n_j)-2|C_{ij}|^2}
{\langle\hat n_i\rangle +\langle\hat n_j\rangle}.
\label{squee}
\end{equation}
We observe, from Eqs.(\ref{var}) and (\ref{corr1}) that the statistics is that
of a chaotic (i.e. thermal) state, as obtained in Ref.\cite{PRA} for the lossless case.  
If the two modes are perfectly number-squeezed, then $\xi_{i,j}=0$, whereas if they are independent 
and coherent, $\xi_{i,j}=1$. As it will be clear in the following sections, it is also 
worth to introduce also the atomic density operator
for the linearized matter-wave field $\hat\Psi(\theta)\approx[\sqrt N + a_1 e^{-i(\theta+\delta\tau)}
+a_2 e^{i(\theta+\delta\tau)}]/\sqrt{2\pi}$, defined as
\begin{equation}
\hat n(\theta)=\hat\Psi^\dag(\theta)\hat\Psi(\theta)\approx\frac{N}{2\pi}
\left(1+\hat B e^{-i(\theta+\delta\tau)}+\hat B^\dag e^{i(\theta+\delta\tau)}\right),
\label{density}
\end{equation}
where $\hat B=(a_1^\dag+a_2)/\sqrt N$ is the bunching operator, with
$\langle\hat B\rangle=0$ and
\begin{equation}
\langle\hat B^\dag\hat B\rangle=\frac{1}{N}(C_{11}+C_{22}+C_{12}+C_{21}).
\label{bunch}
\end{equation}

\section{Analysis of working regimes}\label{s:wr}

We now investigate the different regimes of operation of CARL. For sake of
simplicity, we will discuss only the case with $\gamma_1=\gamma_2=\gamma$, so
that $\gamma_+=\gamma$ and $\gamma_-=0$. In this case the cubic equation
(\ref{cubica}) becomes:
\begin{equation}
\left[\omega-\delta-i\left(\kappa-\gamma\right)\right]
\left(\omega^2-\frac{1}{\rho^2}\right)+1=0
\label{cubica2}
\end{equation}
We will discuss two pairs different regimes of CARL, as defined in ref.\cite{Gatelli}, i.e.:
i) semi-classical good-cavity regime ($\rho\gg 1$ and $\kappa\ll 1$);
ii) quantum good-cavity regime ($\kappa^2\ll\rho<1$);
iii) semi-classical superradiant regime ($\rho\gg\sqrt{2\kappa}>1$);
iv) quantum superradiant regime ($\kappa^2\gg\sqrt{2\kappa}>\rho$).
Also, we  note that the case $\gamma=\kappa$ worth a special
attention. In fact, in this case Eq.(\ref{cubica2}) is independent on losses: 
the effect of decoherence is only a overall factor
$\exp(-\gamma\tau)$ multiplying the functions $f_{ij}$, elements of the matrix
$\mathbf{M}$. Hence, it is expected that the case $\gamma=\kappa$ will have
statistical properties similar to those of the ideal case without losses, as
it will be discussed below.

\subsection{CARL instability}
\begin{figure}[b]
\begin{center}
\includegraphics[width=5.5cm]{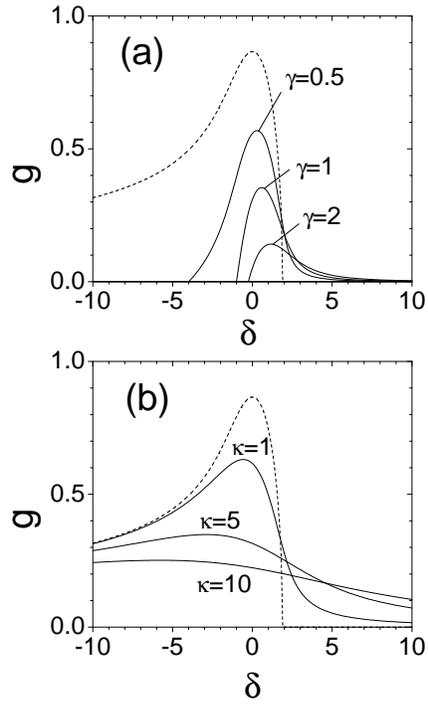}
\caption{Growth rate $g=-{\rm Im}\omega-\gamma$ vs. $\delta$ for the unstable
root of the cubic equation (\ref{cubica}) in the semi-classical limit,
$\rho=100$.  In (a) $\kappa=0$ and $\gamma=0.5,1,2$; in (b) $\gamma=0$ and
$\kappa=1,5,10$. The dashed lines represent the case $\kappa=\gamma=0$.}
\label{fig1}
\end{center}
\end{figure}

First, we investigate the effect of decoherence and cavity losses on the CARL
instability in the different regimes. For large values of $ \tau$ the
functions $f_{ij}$ of Eq.(\ref{eAt}) grow as $\exp(g\tau)$, where $g=-{\rm
Im}\omega-\gamma$ is the exponential gain and $\omega$ is the unstable root of
Eq.(\ref{cubica2}), with negative imaginary part. In fig.\ref{fig1} we plot
$g$  vs.  $\delta$ in the semi-classical regime (e.g. $\rho=100$) for the
good-cavity case ($\kappa=0$) and $\gamma=0.5,1,2$ (fig.\ref{fig1}a), whereas
the transition to the superradiant regime is shown in fig.\ref{fig1}b for
$\kappa=1,5,10$ and $\gamma=0$.  The dashed line in fig.\ref{fig1} shows the
gain $g^{(0)}$ for the ideal case $\kappa=\gamma=0$.  
\par
\begin{figure}
\begin{center}
\includegraphics[width=5.5cm]{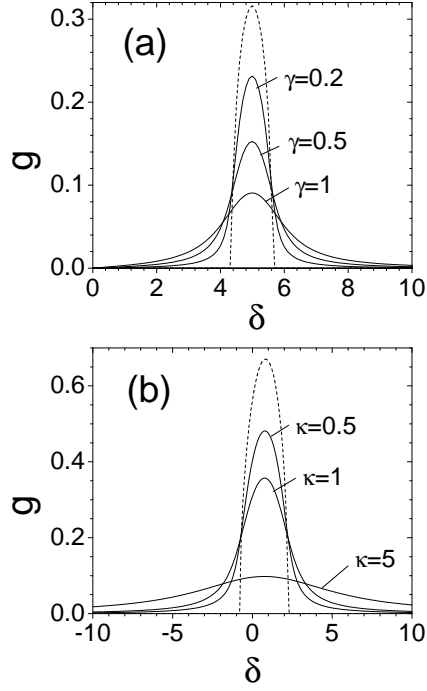}
\caption{Growth rate $g=-{\rm Im}\omega-\gamma$ vs. $\delta$ for the unstable
root of the cubic equation (\ref{cubica}) in the quantum limit. In (a),
$\rho=0.2$, $\kappa=0$ and $\gamma=0.2,0.5,1$; in (b), $\rho=1$, $\gamma=0$
and $\kappa=0.5,1,5$. The dashed lines represent the case $\kappa=\gamma=0$.}
\label{fig2}
\end{center}
\end{figure}

A similar behavior is obtained in the quantum regime shown in fig.\ref{fig2}, where
$g$ is plotted vs. $\delta$ for $\rho=0.2$, $\kappa=0$
and $\gamma=0.2,0.5,1$ (fig.\ref{fig2}a, quantum good-cavity regime) and for
$\rho=1$, $\gamma=0$ and $\kappa=0.5,1,5$ (fig.\ref{fig2}b, quantum
superradiant regime).  Note that, unlike in the semi-classical regime, in
the quantum regime the gain is symmetric around the
resonance $\delta=1/\rho$ (i.e. $\omega_s=\omega_p-\omega_r$).
Notice that in the case $\gamma=\kappa$, $g=g^{(0)}-\gamma$, where $g^{(0)}$ is shown 
by a dashed lines in fig.\ref{fig1} and \ref{fig2}. Whereas increasing
$\kappa$ or $\gamma$ $g$ tends to zero remaining positive for some value of
$\delta$ instead in the case $\gamma=\kappa$ we have a threshold for
$g^{(0)}=\gamma$.

\subsection{Average populations and number squeezing parameter}
\begin{figure}
\begin{center}
\includegraphics[width=5.5cm]{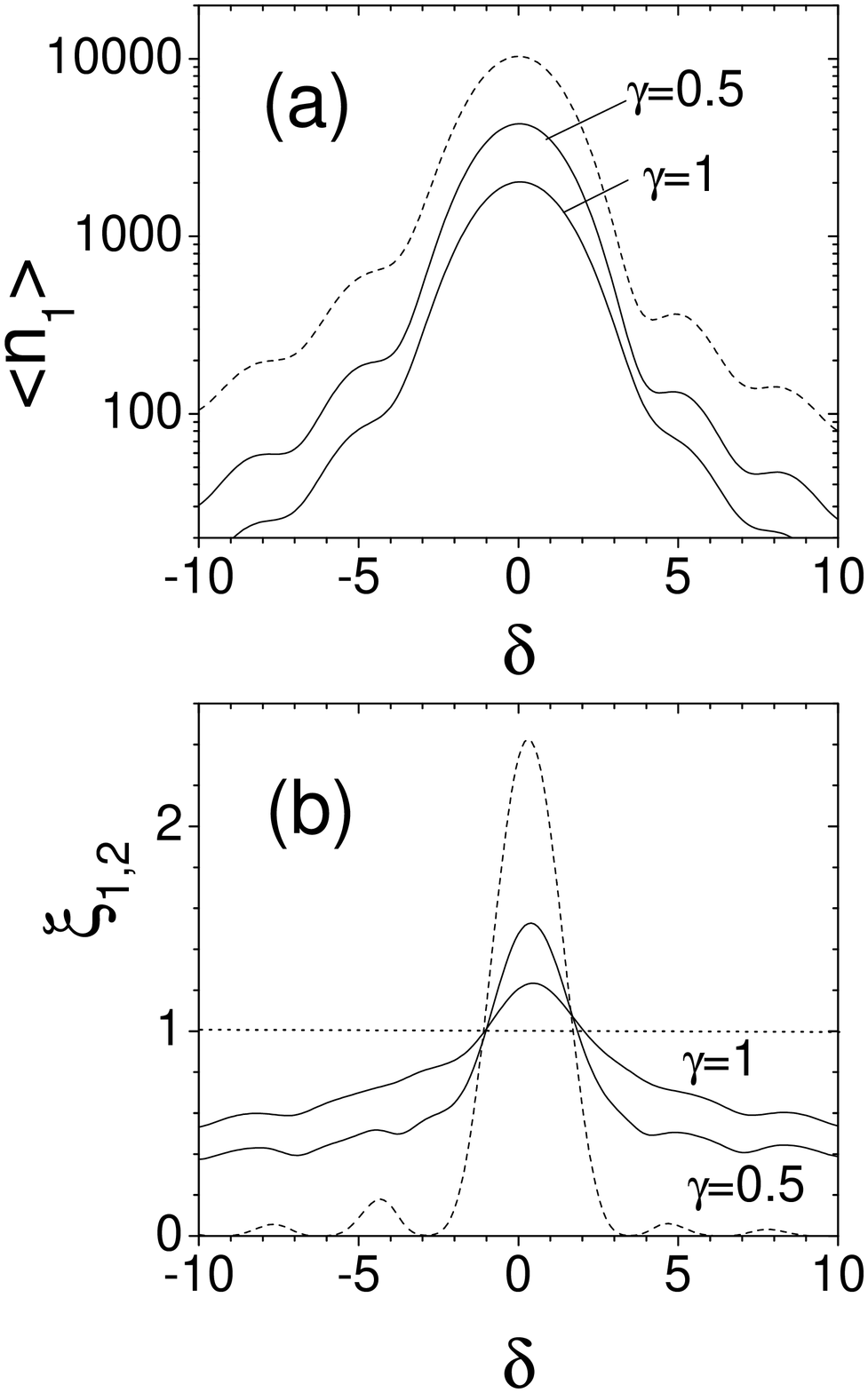}
\caption{Semi-classical regime with $\kappa=0$: $\langle\hat n_1\rangle$, (a),
and $\xi_{1,2}$, (b), vs. $\delta$ for $\rho=100$, $\tau=2$, $\gamma=0$ (dashed
line), $0.5$ and $1$. }
\label{fig3}
\end{center}
\end{figure}

\begin{figure}
\begin{center}
\includegraphics[width=5.5cm]{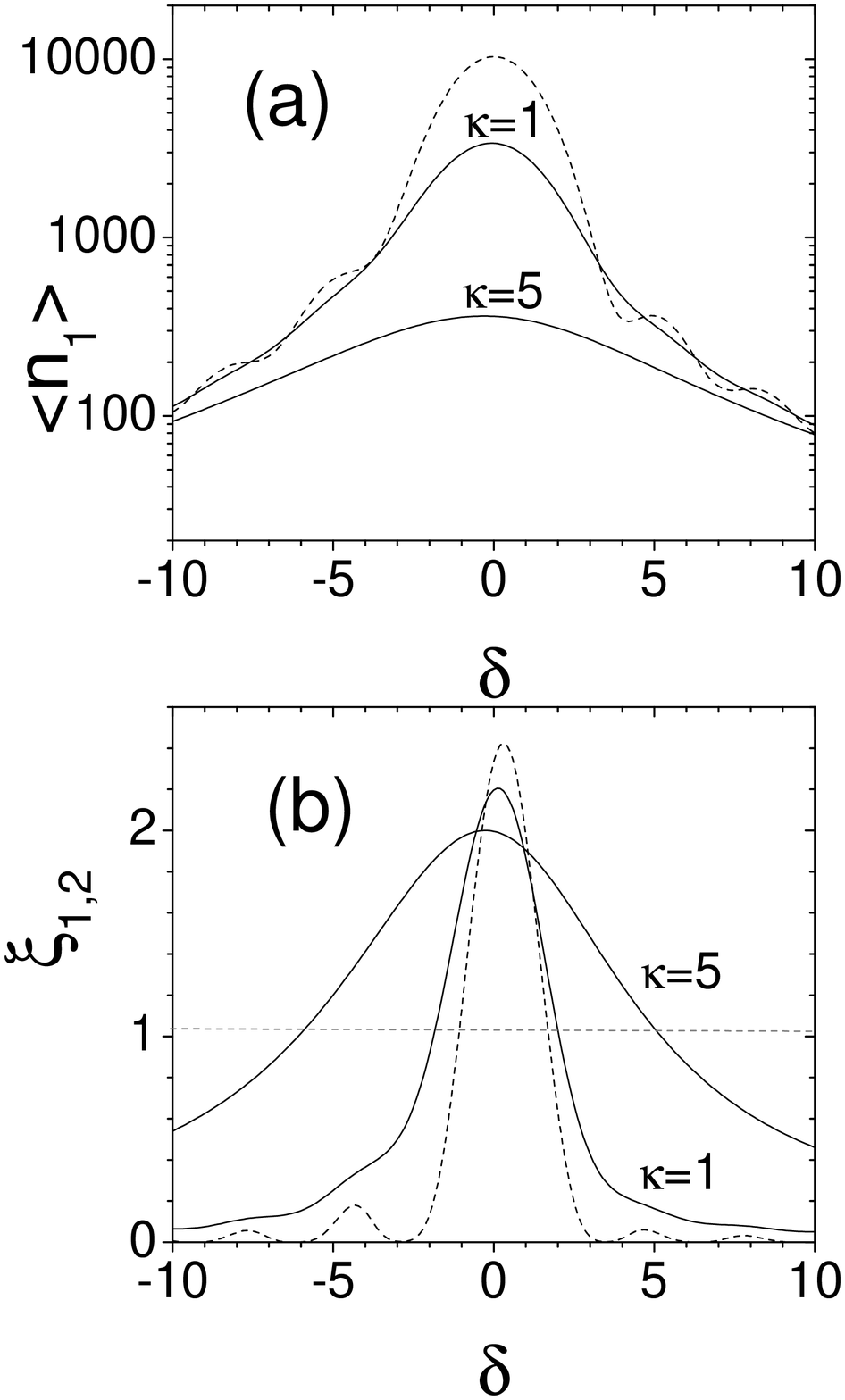}
\caption{Semi-classical regime  with $\gamma=0$: $\langle\hat n_1\rangle$,
(a), and $\xi_{1,2}$, (b), vs. $\delta$ for $\rho=100$, $\tau=2$, $\kappa=0$
(dashed line), $1$ and $5$.}
\label{fig4}
\end{center}
\end{figure}

\begin{figure}
\begin{center}
\includegraphics[width=5.5cm]{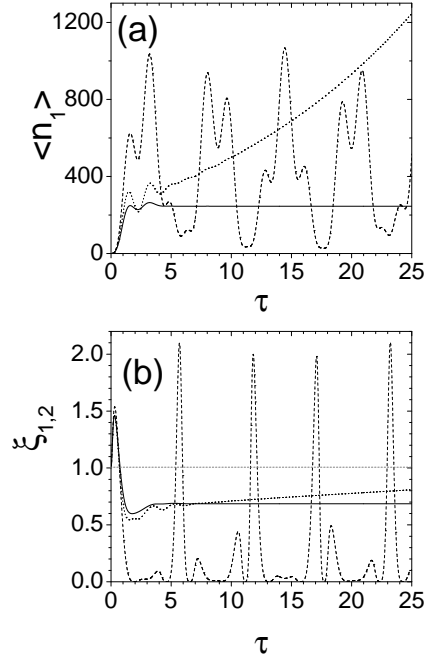}
\caption{Semi-classical regime  for $\rho=100$ and $\delta=3.5$:
$\langle\hat n_1\rangle$, (a), and $\xi_{1,2}$, (b), vs. $\tau$
for $\gamma=0$, $\kappa=0$ (dashed line), for $\gamma=0.2$,
$\kappa=0$ (dotted line) and for $\gamma=\kappa=0.5$ (continuous
line).} \label{fig5}
\end{center}
\end{figure}

Figures \ref{fig3} and \ref{fig4} show the effect of losses, in the
semi-classical regime, on the atomic population $\langle\hat n_1\rangle$, (a),
and on the number squeezing parameter $\xi_{1,2}$, (b), plotted as a function
of $\delta$ for $\rho=100$ and $\tau=2$. Fig.\ref{fig3} shows the effect of
the atomic decoherence on the the high-Q cavity regime ($\kappa=0$) for
$\gamma=0$ (dashed line), $0.5$ and $1$. We observe that increasing $\gamma$
the population of the mode 1 decreases and  the number squeezing parameter
$\xi_{1,2}$ increases in the region of detuning where is less than one, i.e.
where atom-atom number squeezing occurs.  A similar behavior can be observed
increasing the radiation losses in the semi-classical regime, as shown in
fig.\ref{fig4}, where $\langle\hat n_1\rangle$, (a), and $\xi_{1,2}$, (b), are
plotted vs. $\delta$ for $\gamma=0$, $\kappa=0,1,5$ and $\rho=100$. In both
the cases, in order to observe number squeezing in the semi-classical regime,
it is necessary to detune the probe field from resonance, as it was already
pointed out in ref.\cite{PRA}.
The inclusion of losses allows also to reach a steady-state regime when the
gain $g$ is negative.  In this case, the covariance matrix
$\mathbf{C}(\infty)=\mathbf{Q}(\infty)$ becomes asymptotically constant.  An
example of this behavior is shown in fig.\ref{fig5}, where $\langle\hat
n_1\rangle$, (a), and $\xi_{1,2}$, (b), are plotted vs.  $\tau$ for $\rho=100$
and $\delta=3.5$.  The dashed line shows the ideal case $\gamma=\kappa=0$:
because $g^{(0)}=0$ (as it can be observed from fig.\ref{fig1}), the solution
is oscillating and the two atomic modes 1 and 2 are periodically number
squeezed.  The dotted line of fig.\ref{fig5} shows the case with $\kappa=0$
and $\gamma=0.2$. Here, $g=0.025$ and both the average population and the
number squeezing parameter grow in time.  Finally, the continuous line of
fig.\ref{fig5} shows the case $\gamma=\kappa=0.5$: the gain is $g=-0.5$ and
the system reaches a stationary state in which $\xi_{1,2}=0.7$.  This case is
of some interest because a steady-state atom-atom number squeezed state is
obtained in a linear system.
\par
\begin{figure}
\begin{center}
\includegraphics[width=5.5cm]{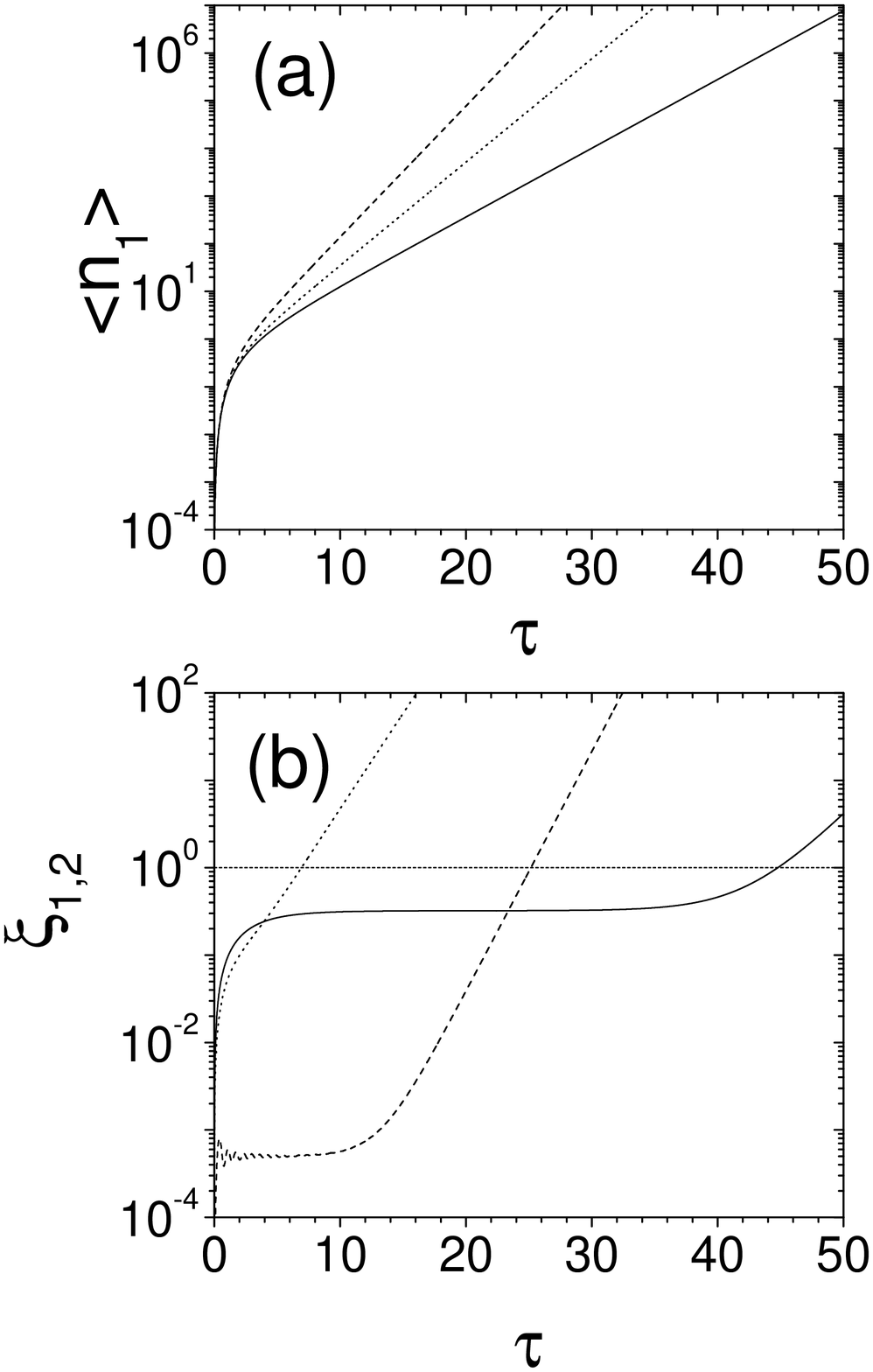}
\caption{Quantum regime  for $1/\rho=\delta=5$: $\langle\hat
n_1\rangle$, (a), and $\xi_{1,3}$, (b), vs. $\tau$  for $\gamma=0$,
$\kappa=0$ (dashed line), for $\gamma=0.15$, $\kappa=0$ (dotted
line) and for $\gamma=\kappa=0.15$ (continuous line).}
\label{fig6}
\end{center}
\end{figure}

Let now consider the effect of losses on the quantum regime.  Fig.\ref{fig6}
shows the average population $\langle\hat n_1\rangle$, (a), and the
atom-photon number squeezing parameter $\xi_{1,3}$, (b), as a function of
$\tau$ for $1/\rho=\delta=5$. Dashed lines in fig.\ref{fig6} a and b are for
$\kappa=\gamma=0$, the dotted lines are for $\kappa=0$ and $\gamma=0.15$ and
the continuous lines are for $\kappa=\gamma=0.15$.  We note that the atomic
decoherence (i.e. $\gamma$) causes a drastic reduction of the
number-squeezing between atoms and photons.  However, choosing
$\gamma=\kappa<g^{(0)}$ (where $g^{(0)}<\sqrt{\rho/2}$), we may keep
$\xi_{1,3}$ constant and less than one for a relatively long time, like in the
ideal case case without losses. Notice that in the quantum regime the
below-threshold regime (i.e. $g<\gamma$) is not of interest because the
average number of quanta generated in each modes remains less than one.

\subsection{Superradiant regime}

In this section we present analytical results for the superradiant regime in
the asymptotic limit $|{\rm Im}\omega_1|\tau\gg 1$, where $\omega_1$ is the
unstable root of Eq.(\ref{cubica2}) with negative imaginary part. For
$\kappa\gg|\omega_1|$ and assuming for simplicity $\gamma=0$, one root of
Eq.(\ref{cubica2}) can be discharged as it decays to zero as
$\exp{(-\kappa\tau)}$ and the other two roots may be obtained solving the
following quadratic equation: 
\begin{equation}
\omega^2+\frac{\omega+\delta+i\kappa}{(\delta+i\kappa)^2-1/\rho^2}-\frac{1}{\rho^2}=0.
\label{quad} \end{equation}
From Eq.(\ref{quad}) it is possible to calculate explicitly the unstable root
and evaluate asymptotically the expressions of the function $f_{ij}$ appearing
in Eq.(\ref{eAt}). From them, it is possible to evaluate the expectation
values of the occupation numbers in the semi-classical and quantum regimes.

\subsubsection{Semi-classical limit of the superradiant regime}
For $\kappa^{3/2}> 1\gg \sqrt{\kappa}/\rho$ and $\delta=0$, the solutions of
Eq.(\ref{quad}) are $\omega_{1,2}=\approx(1\mp i)/\sqrt{2\kappa}$ and the
average occupation numbers are: 
\begin{eqnarray} \langle
n_1\rangle&\approx&\frac{\rho^2}{16\kappa}
\left[1+\frac{\sqrt{2\kappa}}{\rho}\right]e^{(2/\kappa)^{1/2}\tau}
\label{ave:sr:1}\\ \langle
n_2\rangle&\approx&\frac{\rho^2}{16\kappa}e^{(2/\kappa)^{1/2}\tau}
\label{ave:sr:2}\\ \langle
n_3\rangle&\approx&\frac{\rho}{8\kappa^2}e^{(2/\kappa)^{1/2}\tau}\label{ave:sr:3}
\end{eqnarray} 
We observe that $\langle n_1\rangle\approx\langle n_2\rangle$
and $\langle n_3\rangle\approx(2/\rho\kappa)\langle n_1\rangle\ll \langle
n_1\rangle$, so that the number of emitted photons is much smaller than the
number of atoms in the two motional states. The asymptotic expression of the
expectation value (\ref{bunch}) of the bunching parameter is 
\begin{equation} \langle\hat
B^\dag\hat
B\rangle\approx\frac{1}{4N}\left[1+\frac{\sqrt{2\kappa}}{\rho}\right]e^{(2/\kappa)^{1/2}\tau}.
\label{bunch:sc} \end{equation} Assuming that $\langle\hat B^\dag\hat
B\rangle$ approaches a maximum value of the order of one, then the maximum
average number of emitted photons  is about $\rho N/2\kappa^2$, whereas the
maximum fraction of atoms gaining a momentum $2\hbar k_p$ is about
$\rho^2/4\kappa$.

\subsubsection{Quantum limit of the superradiant regime}
For $\kappa^{3/2}\gg 1> \rho/\sqrt{\kappa}$ and $\delta=1/\rho$, the solutions
of Eq.(\ref{quad}) are $\omega_{1,2}\approx 1/\rho\mp i\rho/(2\kappa)$ and the
average occupation numbers are:
\begin{eqnarray}
\langle
n_1\rangle&\approx&\left[1+\left(\frac{\rho}{\sqrt{2\kappa}}\right)^4\right]e^{(\rho/\kappa)\tau}
\label{ave:srq:1}\\ \langle
n_2\rangle&\approx&\left(\frac{\rho}{2\sqrt{\kappa}}\right)^4
e^{(\rho/\kappa)\tau}\label{ave:srq:2}\\ \langle
n_3\rangle&\approx&\frac{\rho}{2\kappa^2}e^{(\rho/\kappa)\tau}\label{ave:srq:3}.
\end{eqnarray}
In this case, $\langle n_{2,3}\rangle\ll\langle n_1\rangle$ and  $\langle
n_2\rangle\approx(\rho/2)^3\langle n_3\rangle$: the average number of emitted
photons is much less than the average number of atoms scattering a photon from
the pump to the probe.  Furthermore, the number of atoms making the reverse
process, i.e. scattering a photon from the probe to the pump, can be larger
than the number of photons scattered into the probe mode if $\rho>2$, as it
occurs in the current experiment on BEC superradiance \cite{MIT,LENS}. In this
regime the asymptotic expression of the expectation value of the bunching parameter
is:
\begin{equation}
\langle\hat B^\dag\hat B\rangle\approx\frac{1}{N}\left[1+
\frac{1}{2}\left(\frac{\rho}{\sqrt{2\kappa}}\right)^4\right]e^{(\rho/\kappa)\tau},
\label{bunch:q}
\end{equation}
so that  $\langle n_{3}\rangle\approx (\rho N/2\kappa^2)\langle\hat B^\dag\hat
B\rangle$, as in the semi-classical limit.  The only difference is that in the
quantum regime the maximum of $\langle\hat B^\dag\hat B\rangle$ is $1/2$, so
that the maximum number of scattered photon in the quantum limit is half of
that obtained in the semi-classical limit.

\section{Entanglement and separability}\label{s:en}
In this Section we analyze the kind of entanglement that
can be generated from our system. First, we establish notation and
illustrate the separability criteria. We also apply the criteria
to the state obtained in the ideal dynamics. Then,
we address the effects of losses. We study both the
separability  properties of the tripartite state resulting from
the evolution from the vacuum, as well as of the three two-mode
states that are obtained by partial tracing over one of the modes.
The basis of our analysis is that both the tripartite state and
the partial traces are Gaussian states at any time. Therefore, we
are able to fully characterize three-mode and two-mode
entanglement as a function of the interaction  parameters
\cite{simon,Giedke}.

\subsection{Three-mode entanglement}\label{ss:3mode}
Concerning entanglement properties, three-mode states may be
classified as follows \cite{Giedke}:
\begin{itemize}
\item[{\em Class 1}]: fully inseparable states, i.e. not separable for any grouping of
the modes;
\item[{\em Class 2}]: one-mode biseparable states, which are separable if two of the modes
are grouped together, but inseparable with respect to the other
groupings;
\item[{\em Class 3}]: two-mode biseparable states, which are separable with respect to
two of the three possible bipartite groupings but inseparable with
respect to the third;
\item[{\em Class 4}]: three-mode biseparable states, which are separable with respect to
all three bipartite groupings, but cannot be written as a product
state;
\item[{\em Class 5}]: fully separable states, which can be written as a three-mode
product state.
\end{itemize}
Separability properties are determined by the characteristic
function. In order to simplify the analysis we rewrite the
characteristic function (\ref{chfun}) in terms of the real
variables ${\mathbf x}^T\equiv (x_1,x_2,x_3,y_1,y_2,y_3)$ with
$\xi_j= 2^{-1/2}(y_j-i x_j)$, $j=1,2,3$. We have
\begin{equation}
\chi({\mathbf x}) = \exp \left\{ - \frac14 {\mathbf x}^T {\mathbf
V}{\mathbf x} \right\} \label{c6}\;,
\end{equation}
where
\begin{equation}
{\mathbf V} = 2 {\mathbf \Lambda}_0 \left(\begin{array}{cc}
{\mathbf A}& -{\mathbf B } \\
{\mathbf B}& {\mathbf A } \end{array}\right)\: {\mathbf \Lambda}_0
\label{defV}\;,
\end{equation}
with ${\mathbf \Lambda}_0=\hbox{Diag}(-1,1,1,1,1,1)$ and
\begin{equation}
{\mathbf A}= \Re {\it e}\:{\mathbf C }  \qquad {\mathbf B}= \Im
{\it m}\: {\mathbf C } \label{defAB}\;,
\end{equation}
and where we omitted the explicit time dependence of the matrices.
The entanglement properties of the three-mode state are determined
by the positivity of the matrices $$\Gamma_j=\Lambda_j {\mathbf V}
\Lambda_j - i {\mathbf J}\qquad j=1,2,3$$ 
where
$\Lambda_1=\hbox{Diag}(1,1,1,-1,1,1)$,
$\Lambda_2=\hbox{Diag}(1,1,1,1,-1,1)$,
$\Lambda_3=\hbox{Diag}(1,1,1,1,1,-1)$ and ${\mathbf J}$ is the
symplectic block matrix
\begin{eqnarray}
{\mathbf J}=\left(\begin{array}{cc}0  &-{\mathbf I} \\ {\mathbf
I} &\quad 0\end{array}\right)\:, \label{matJ}
\end{eqnarray}
${\mathbf I}$ being the $3\times3$ identity matrix. The positivity
of the matrix $\Gamma_j$ indicates that the $j$-th mode may be
factorized from the other two. Therefore, we have that i) if
$\Gamma_j< 0$ $\forall j$ the state is in class 1; ii) if only one
of the $\Gamma_j$ is positive the state is in class 2; iii) is
only two of the $\Gamma_j$ are positive the state is in class 3;
iv) if $\Gamma_j >0$, $\forall j$ then the state is either in
class 4 or in class 5.
\par
\begin{figure}
\begin{center}
\includegraphics[width=5.5cm]{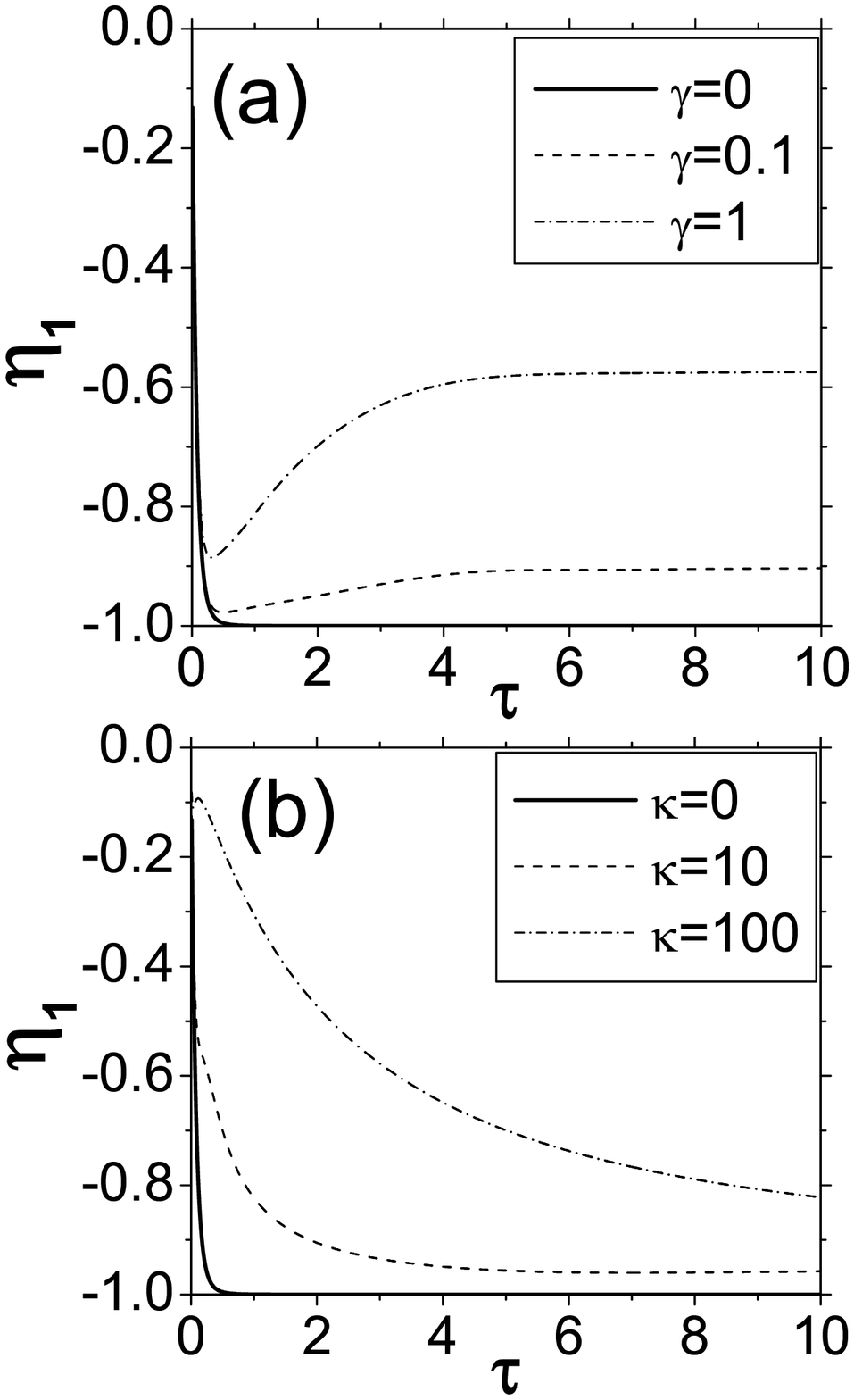}
\caption{Semi-classical regime for $\rho=100$ and $\delta=0.01$:
minimum eigenvalue of matrix $\Gamma_1$ for $\kappa=0$ and
different values of $\gamma$ (a) and for $\gamma=0$ and
different values of $\kappa$ (b).} \label{fig7}
\end{center}
\end{figure}
\begin{figure}
\begin{center}
\includegraphics[width=5.5cm]{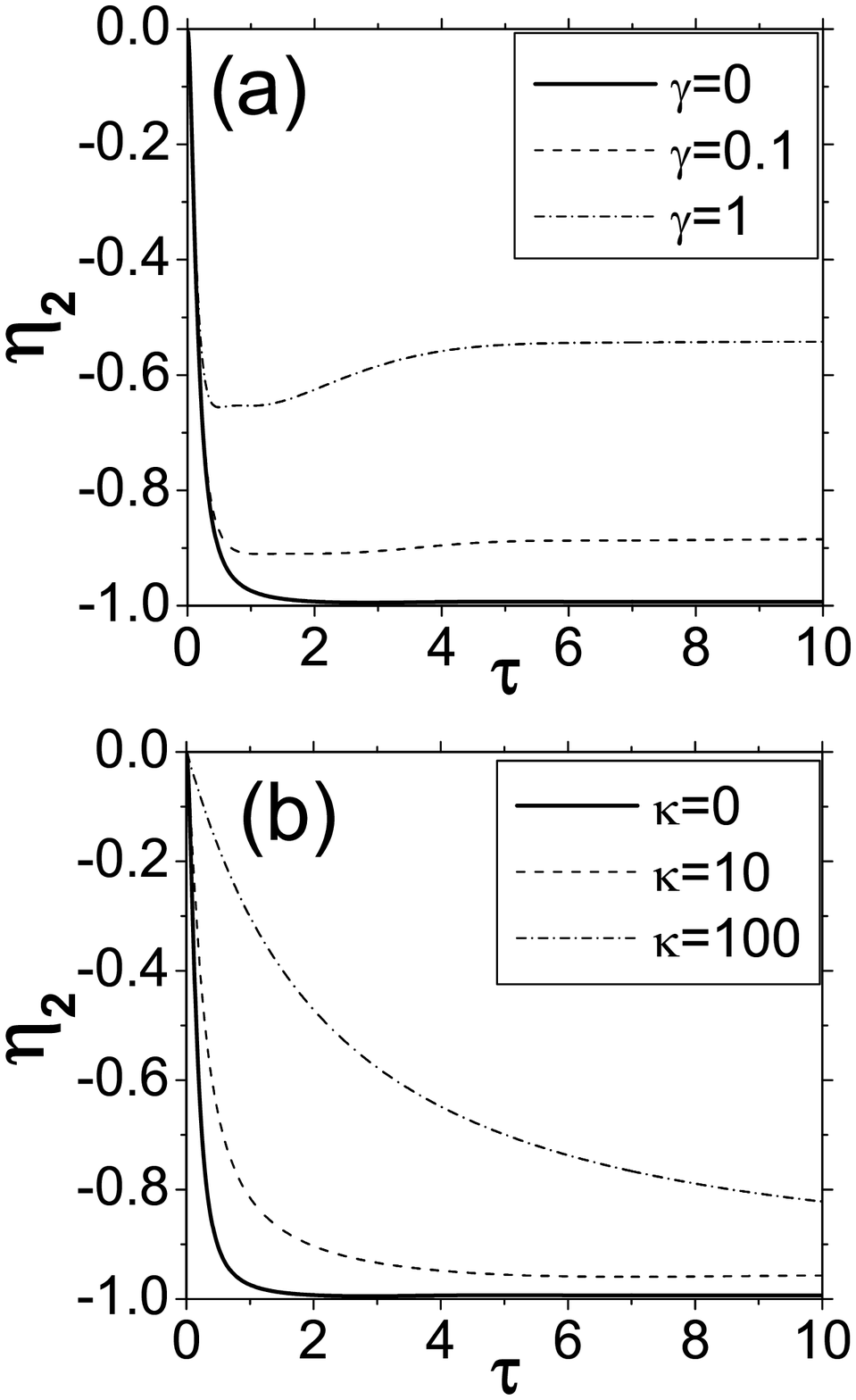}
\caption{Semi-classical regime for $\rho=100$ and $\delta=0.01$:
minimum eigenvalue of matrix $\Gamma_2$ for $\kappa=0$ and
different values of $\gamma$ (a) and for $\gamma=0$ and
different values of $\kappa$ (b).} \label{fig8}
\end{center}
\end{figure}
\begin{figure}
\begin{center}
\includegraphics[width=5.5cm]{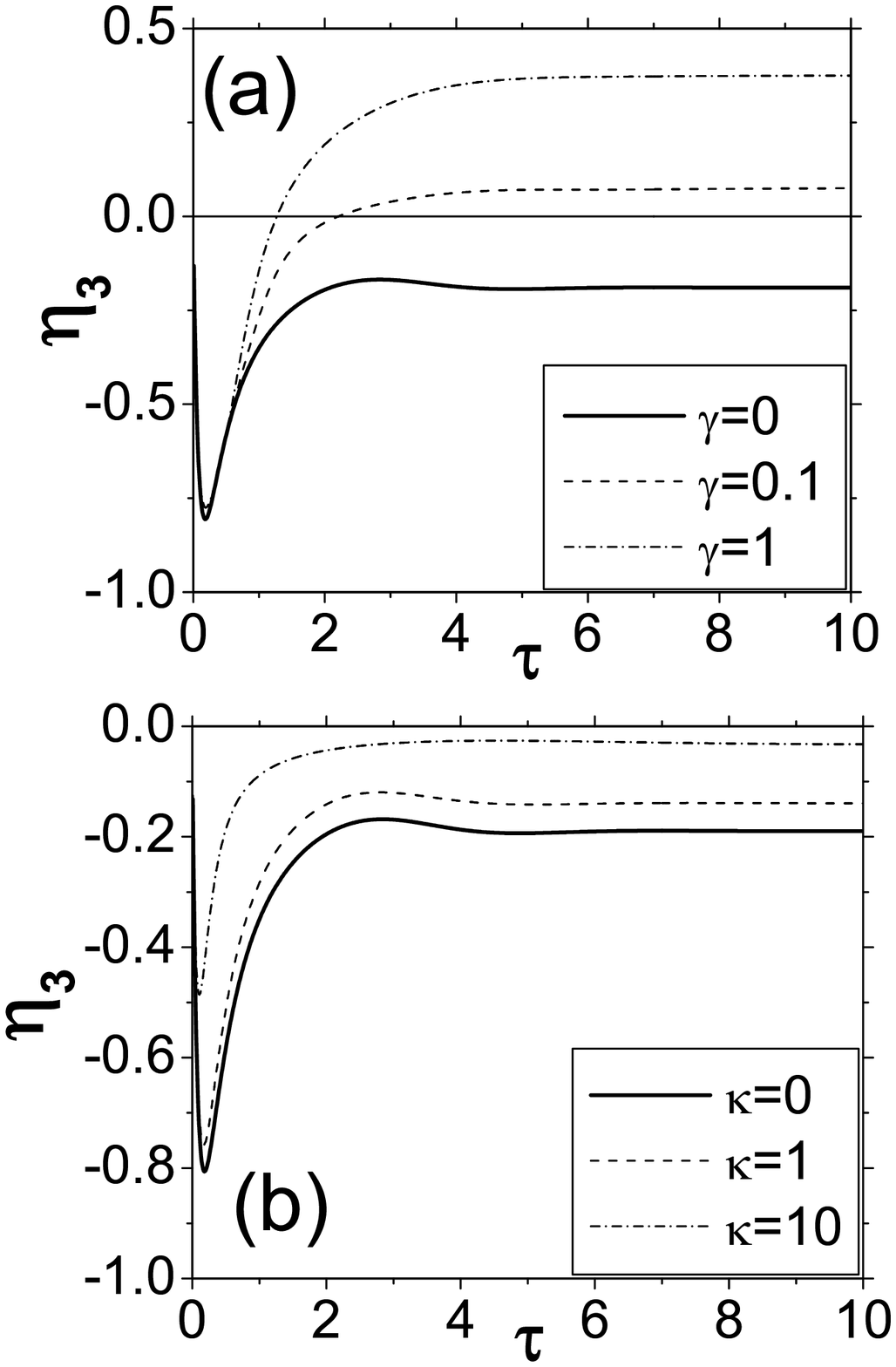}
\caption{Semi-classical regime for $\rho=100$ and $\delta=0.01$:
minimum eigenvalue of matrix $\Gamma_3$ for $\kappa=0$ and
different values of $\gamma$ (a) and for $\gamma=0$ and
different values of $\kappa$ (b).} \label{fig9}
\end{center}
\end{figure}
\begin{figure}
\begin{center}
\includegraphics[width=5.5cm]{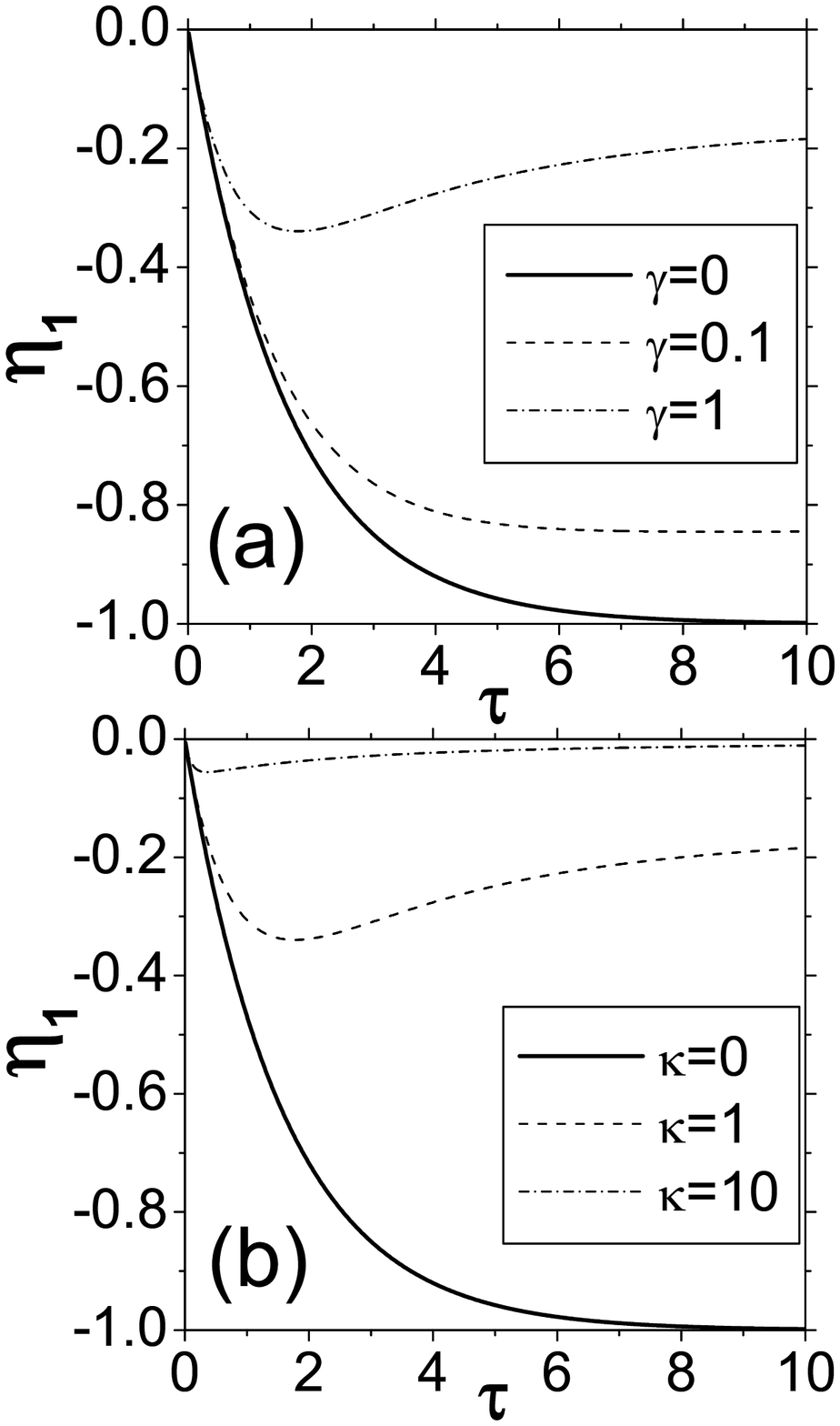}
\caption{Quantum regime for $\rho=0.2$ and $\delta=5$: minimum
eigenvalue of matrix $\Gamma_1$ (or $\Gamma_3$, see the text) for $\kappa=0$ and different
values of $\gamma$ (a) and for $\gamma=0$ and different values
of $\kappa$ (b).} \label{fig10}
\end{center}
\end{figure}
\begin{figure}
\begin{center}
\includegraphics[width=5.5cm]{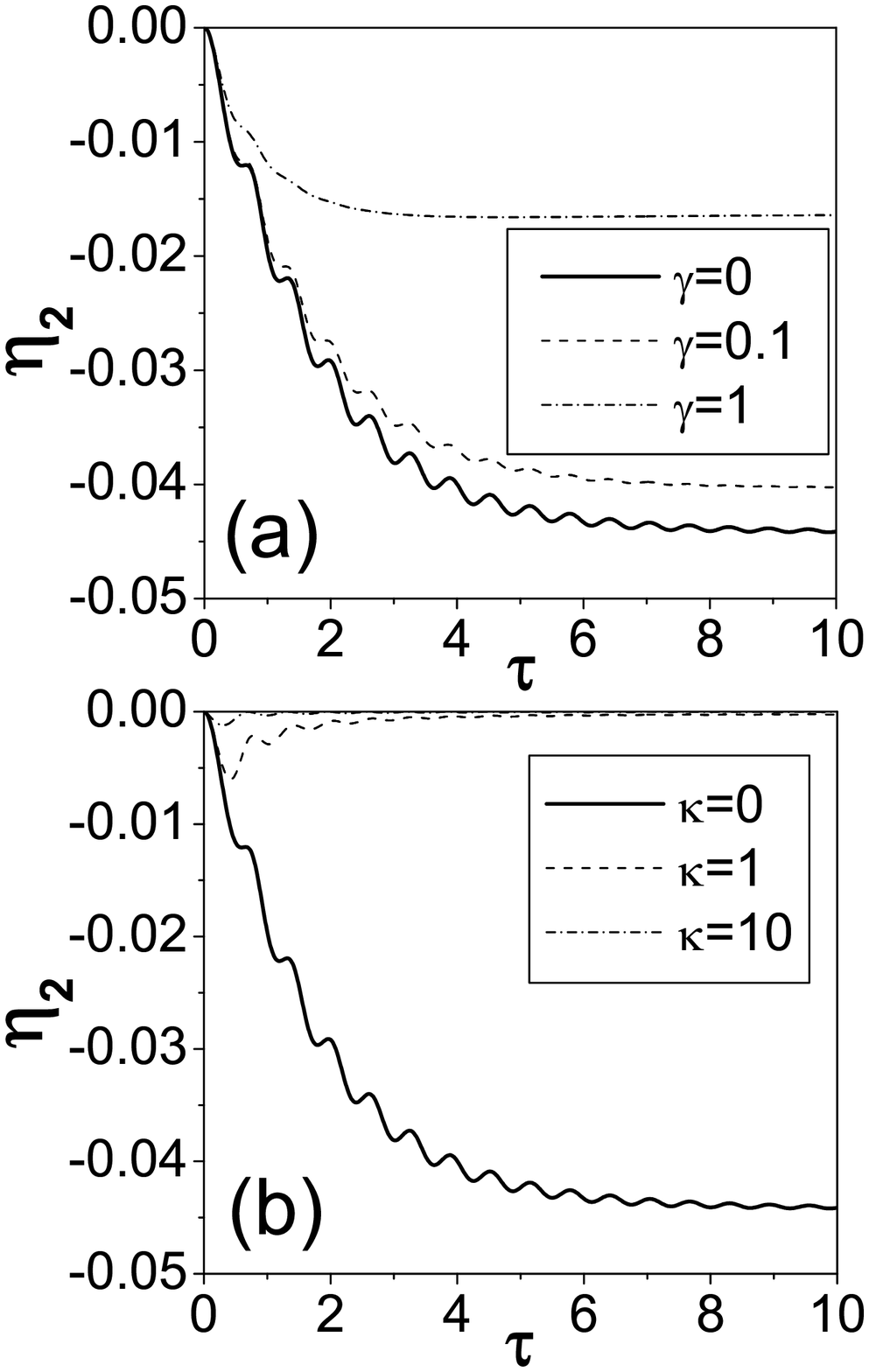}
\caption{Quantum regime for $\rho=0.2$ and $\delta=5$: minimum
eigenvalue of matrix $\Gamma_2$ for $\kappa=0$ and different
values of $\gamma$ (a) and for $\gamma=0$ and different values
of $\kappa$ (b).} \label{fig11}
\end{center}
\end{figure}

The covariance matrix ${\mathbf V}$ can be written as
\begin{equation}
{\mathbf V} = \left(
\begin{array}{cccccc}
{\cal G} & -{\cal A} & - {\cal B} & 0 & {\cal D} & {\cal E} \\
- {\cal A}  &{\cal H}&  {\cal C}  & {\cal D} & 0 & - {\cal F} \\
- {\cal B} &  {\cal C} & {\cal I}  & {\cal E} & {\cal F} & 0        \\
0           & {\cal D} & {\cal E} & {\cal G}& {\cal A} & {\cal B} \\
{\cal D} &0          & {\cal F} & {\cal A} & {\cal H} & {\cal C} \\
{\cal E} & -{\cal F} &0           & {\cal B}& {\cal C} & {\cal I} \\
\end{array}
\right) \label{matV}\;,
\end{equation}
where
\begin{eqnarray}
{\cal A} = 2 \: \Re{\it e}\: C_{12} \quad & {\cal D} = 2 \:
\Im{\it m}\: C_{12} \quad& {\cal G} = 2\langle\hat n_1\rangle+1
\nonumber \\
{\cal B} = 2 \: \Re{\it e}\: C_{13}\quad & {\cal E} = 2 \: \Im{\it
m}\: C_{13} \quad& {\cal H} = 2\langle\hat n_2\rangle+1
 \label{defcalquant} \\
{\cal C} = 2 \: \Re{\it e}\: C_{23}\quad & {\cal F} = 2 \: \Im{\it
m}\: C_{23} \quad& {\cal I} = 2\langle\hat n_3\rangle+1 \nonumber \;
\end{eqnarray}
and the matrix elements $C_{ij}$ are reported in Appendix
\ref{a:appB}. Let us first consider the ideal case, when no losses
are present. In this case we can prove analytically that the
evolved state (\ref{Tstate}) is fully inseparable. In fact, we
have that
\begin{eqnarray}
{\cal A} =2 \sqrt{\langle\hat n_2\rangle(1+\langle\hat n_1\rangle)} \cos\phi_2 \quad 
& {\cal D} = 2 \sqrt{\langle\hat n_2\rangle(1+\langle\hat n_1\rangle)} \sin\phi_2
\nonumber \\
{\cal B} = 2 \sqrt{\langle\hat n_3\rangle(1+\langle\hat n_1\rangle)} \cos\phi_3 \quad
& {\cal E} = 2 \sqrt{\langle\hat n_3\rangle(1+\langle\hat n_1\rangle)} \sin\phi_3
\nonumber \\
{\cal C} =2 \sqrt{\langle\hat n_2\rangle\langle\hat n_3\rangle}
\cos(\phi_3-\phi_2) \quad 
& {\cal F} =2 \sqrt{\langle\hat n_2\rangle\langle\hat n_3\rangle} \sin(\phi_3-\phi_2) \;,
\end{eqnarray}
from which, in turn, it is straightforward to prove that the
minimum eigenvalues of the matrices $\Gamma_j$ are always
negative.
In the non ideal case, when $\gamma$ or $\kappa$ are different
from zero, the expressions given in Eqs. (\ref{defcalquant}) and
accordingly the minimum eigenvalues of matrices $\Gamma_j$ should be
calculated numerically. In Fig. \ref{fig7}, \ref{fig8} and
\ref{fig9} the minimum eigenvalues of matrices $\Gamma_1$, $\Gamma_2$ and
$\Gamma_3$ are plotted  in the semi-classical regime, with $\rho=100$. In this regime we
can observe that modes $1$ and $2$ remain non separable from the
three mode state even for large values of atomic decoherence
$\gamma$ and radiation losses $\kappa$. Instead inseparability of
mode $3$ is not so robust especially in presence of some atomic
decoherence. In Fig. \ref{fig10} and \ref{fig11} the
minimum eigenvalues of matrices $\Gamma_1$ and $\Gamma_2$ are plotted
in the quantum regime, with $\rho=0.2$. 
The minimum eigenvalue of the matrix $\Gamma_3$ is not reported in the figure
since the behavior is similar to that of $\Gamma_1$.  In this regime we can
observe that modes $1$ and $3$ remain non separable from the three mode state
even for large values of atomic decoherence $\gamma$ and radiation losses
$\kappa$. Instead inseparability of mode $2$ is very sensible especially in
presence of some radiation losses. In any case in the quantum regime the three
eigenvalues increasing $\gamma$ and $\kappa$ approaches to zero but remain
negative. In the semi-classical regime the eigenvalue of $\Gamma_3$ that
corresponds to photonic mode $3$ becomes positive increasing $\gamma$.

\subsection{Two-mode entanglement}\label{ss:2mode}
In experimental conditions where only two of the modes are
available for investigations, the relevant piece of information is
contained in the partial traces of the global three-mode state.
Therefore, besides the study of three-mode entanglement it is also
of interest to analyze the two-mode entanglement properties of
partial traces. At first we notice that the Gaussian character of
the state is preserved by the partial trace operation. Moreover,
the covariance matrices $V_{ij}$ of three possible partial traces
$\hat\varrho_{ij}=\hbox{Tr}_k[\hat\varrho]$, $i\neq j\neq k$ can
be obtained from ${\mathbf V}$ by deleting the corresponding
$k$-th and $k+3$-th rows and columns. The Gaussian character of
the partial traces also permits to check separability using the
necessary and sufficient conditions introduced in Ref.
\cite{simon}, namely by the positivity of the matrices ${\mathbf
S}_{ij}$ and ${\mathbf S}_{ij}^{\prime}$ that are obtained by
deleting the $k$-th and $k+3$-th  rows and columns either from
$\Gamma_i$ or $\Gamma_j$. Since they differ only for the sign of
some off-diagonal elements it is easy to prove that they have the
same eigenvalues. Therefore, we employ only ${\mathbf S}_{ij}$ in
checking separability. The matrices ${\mathbf S}_{ij}$ are given
by
\begin{eqnarray}
{\mathbf S}_{12} &=& \left(\begin{array}{cccc}
{\cal G} & - {\cal A} & i & {\cal D} \\
- {\cal A} & {\cal H}&  -{\cal D} & i\\
-i & - {\cal D} & {\cal G} &  -{\cal A}  \\
{\cal D} & - i & - {\cal A} & {\cal H}
\end{array}
\right) \\
{\mathbf S}_{13} &=& \left(\begin{array}{cccc}
{\cal G} & - {\cal B} & i & {\cal E} \\
- {\cal B} & {\cal L}&  -{\cal E} & i\\
-i & - {\cal E} & {\cal G} &  -{\cal B}  \\
{\cal E} & - i & - {\cal B} & {\cal I}
\end{array}
\right)\\
{\mathbf S}_{23} &=& \left(\begin{array}{cccc}
{\cal H} & {\cal C} & i & -{\cal F} \\
{\cal C} & {\cal L}&  -{\cal L} & i \\
-i & - {\cal L} & {\cal H}&  -{\cal C}  \\
-{\cal F} & - i & - {\cal C} & {\cal I}
\end{array}
\right) \label{RedGam}\;.
\end{eqnarray}
In ideal conditions with $\gamma=\kappa=0$ the minimum eigenvalues of ${\mathbf S}_{1k}$,
$k=2,3$ are given by
\begin{equation}
\eta_{1k}= \langle\hat n_1\rangle + \langle\hat n_k\rangle -
\sqrt{4 \langle\hat n_k\rangle+(\langle\hat n_1\rangle +
\langle\hat n_k\rangle)^2}
\end{equation}
and thus are always negative. On the contrary, the minimum
eigenvalue of $S_{23}$ is given by
\begin{equation}
\eta_{23}= 1+ \langle\hat n_1\rangle + \sqrt{(1+\langle\hat
n_1\rangle)^2 - 4 \langle\hat n_k\rangle}
\end{equation}
where $\langle\hat n_k\rangle={\rm max}(\langle\hat n_2\rangle,\langle\hat n_3\rangle)$. Note 
that $\eta_{23}$ is always positive. Therefore, after partial
tracing we may have atom-atom entanglement (entanglement between
mode $a_1$ and mode $a_2$) or scattered atom-radiation
entanglement (entanglement between mode $a_1$ and mode $a_3$) but
no entanglement between mode $a_2$ and mode $a_3$.

For $\tau\gg 1$ we know the asymptotic expressions for
populations in the ideal case without losses \cite{PRA}, so we can obtain the stationary value of
$\eta_{1k}$ as
\begin{equation}
\eta_{1k}\approx -\frac{2\langle\hat n_k\rangle}{\langle\hat n_1\rangle
+ \langle\hat n_k\rangle}.
\end{equation}

\begin{figure}
\begin{center}
\includegraphics[width=5.5cm]{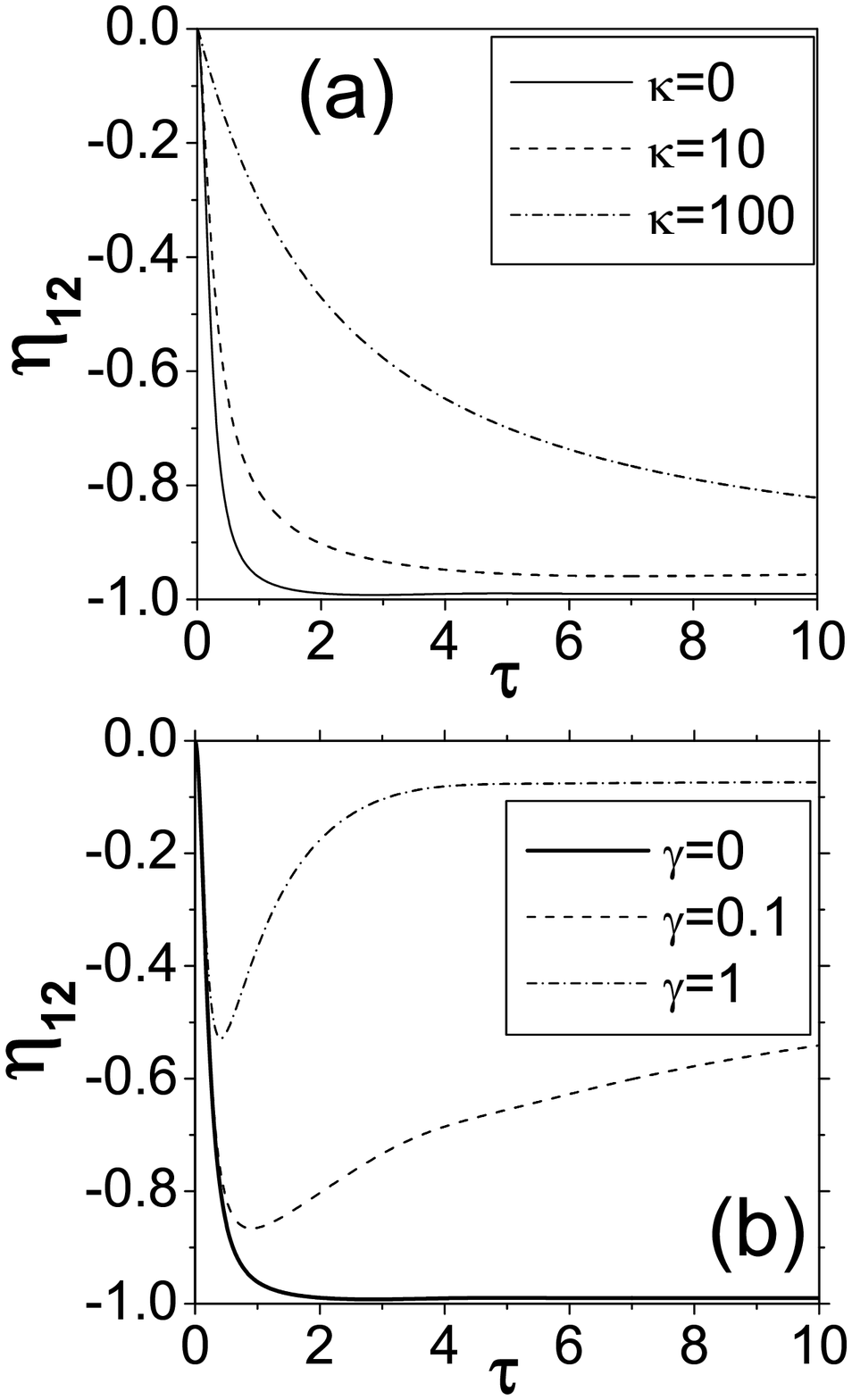}
\caption{Semi-classical regime  for $\rho=100$ and $\delta=0$:
minimum eigenvalue of matrix $S_{12}$ for $\kappa=0$ and different
values of $\gamma$ (a) and for $\gamma=0$ and different values
of $\kappa$ (b).} \label{fig12}
\end{center}
\end{figure}

\begin{figure}
\begin{center}
\includegraphics[width=5.5cm]{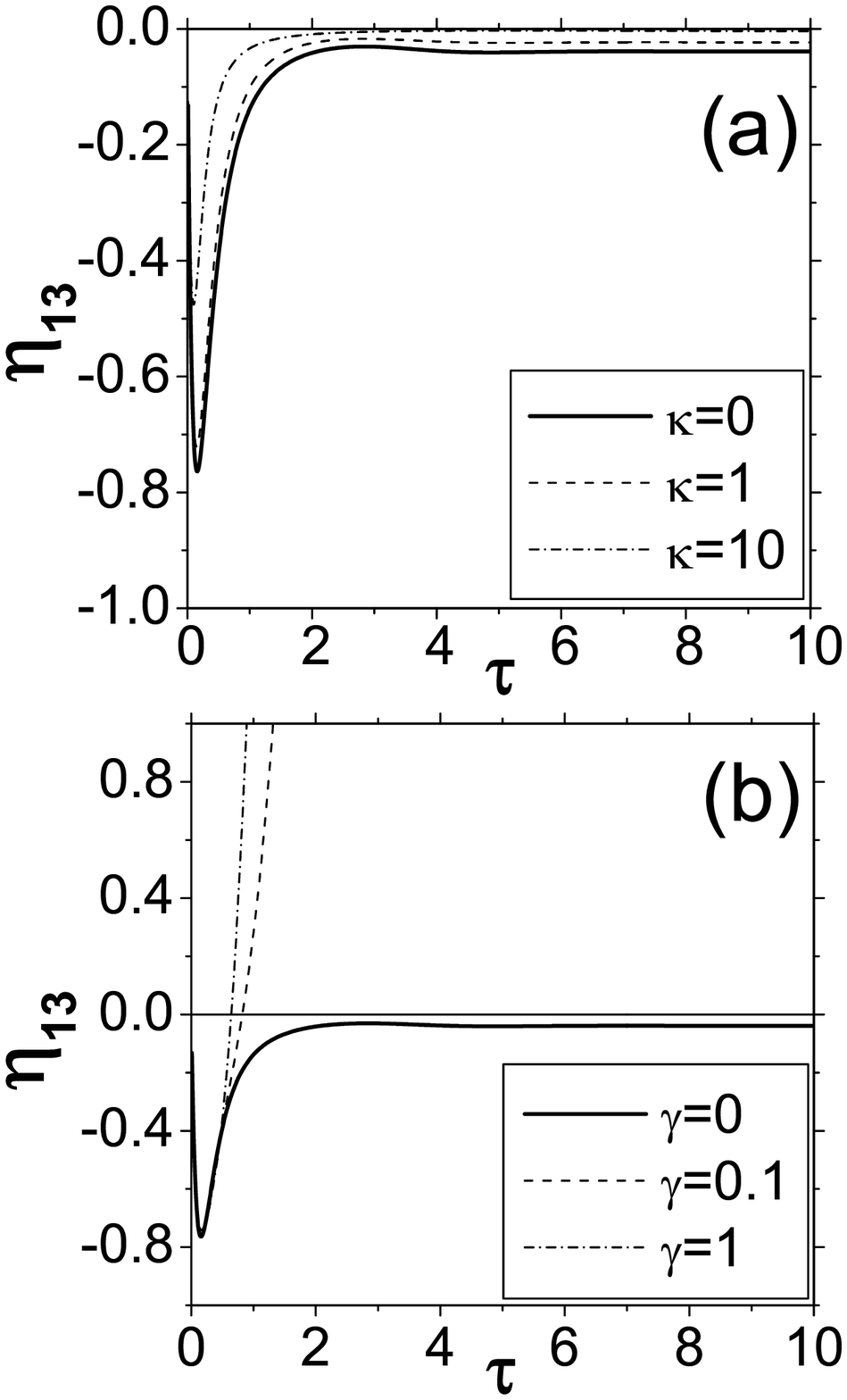}
\caption{Semi-classical regime  for $\rho=100$ and $\delta=0$:
minimum eigenvalue of matrix $S_{13}$ for $\kappa=0$ and different
values of $\gamma$ (a) and for $\gamma=0$ and different values
of $\kappa$ (b).} \label{fig13}
\end{center}
\end{figure}

In the high-gain semi-classical regime ($\rho\gg 1$) \cite{PRA},
\begin{eqnarray}
\langle
\hat{n}_1\rangle&\approx&\frac{1}{18}\left[\frac{\rho^2}{2}+\rho\right]
e^{\sqrt 3\tau},\label{n1:c}\\
\langle \hat{n}_2\rangle&\approx&\frac{\rho^2}{36}
e^{\sqrt 3\tau},\label{n2:c}\\
\langle \hat{n}_3\rangle&\approx&\frac{\rho}{18}e^{\sqrt
3\tau},\label{n3:c}
\end{eqnarray}
so that
\begin{equation}
\eta_{12}\approx -\frac{\rho}{1+\rho} \qquad \eta_{13}\approx -\frac{4}{4+\rho}
\end{equation}
In the high-gain quantum regime ($\rho<1$),
\begin{eqnarray}
\langle n_1\rangle&\approx&\frac{1}{4}\left[1+\left(\frac{\rho}{2}\right)^3\right]
e^{\sqrt{2\rho}\tau},\label{n1:q}\\
\langle
n_2\rangle&\approx&\frac{1}{4}\left(\frac{\rho}{2}\right)^3
e^{\sqrt{2\rho}\tau},\label{n2:q}\\
\langle
n_3\rangle&\approx&\frac{1}{4}e^{\sqrt{2\rho}\tau},\label{n3:q}
\end{eqnarray}
so that
\begin{equation}
\eta_{12}\approx -\frac{\rho^3}{4+\rho^3} 
\qquad \eta_{13}\approx -\frac{16}{16+\rho^3}.
\end{equation}
In the non ideal case, when $\gamma$ or $\kappa$ are different
from zero, the minimum eigenvalues of matrices $S_{12}$ and
$S_{13}$ can be easily obtained
numerically. In Fig. \ref{fig12} and \ref{fig13} the
minimum eigenvalues of matrices $S_{12}$, $S_{13}$ are plotted 
for the semi-classical regime. We can observe that the atom-atom entanglement 
of the reduced state $12$ is robust, as the minimum eigenvalue remain negative 
increasing atomic decoherence $\gamma$ and radiation losses $\kappa$. On the
contrary atom-photon entanglement of the reduced state
$13$ is more sensitive to noise: the eigenvalue remains negative 
increasing $\kappa$ and become positive in presence of some atomic
decoherence.
\par
\begin{figure}
\begin{center}
\includegraphics[width=5.5cm]{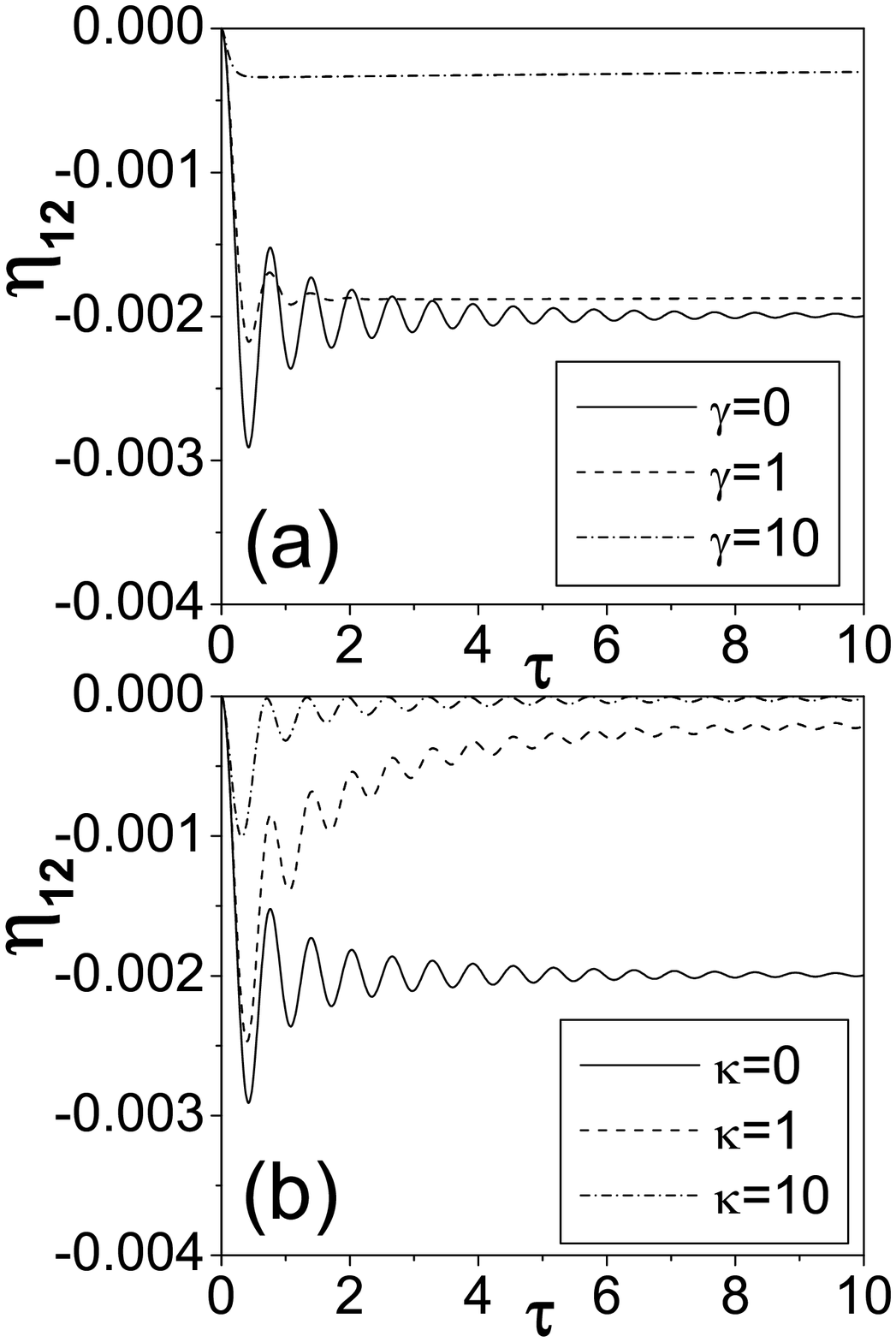}
\caption{Quantum regime  for $\rho=0.2$ and $\delta=5$: minimum
eigenvalue of matrix $S_{12}$ for $\kappa=0$ and different values
of $\gamma$ (a) and for $\gamma=0$ and different values of
$\kappa$ (b).} \label{fig14}
\end{center}
\end{figure}
\begin{figure}
\begin{center}
\includegraphics[width=5.5cm]{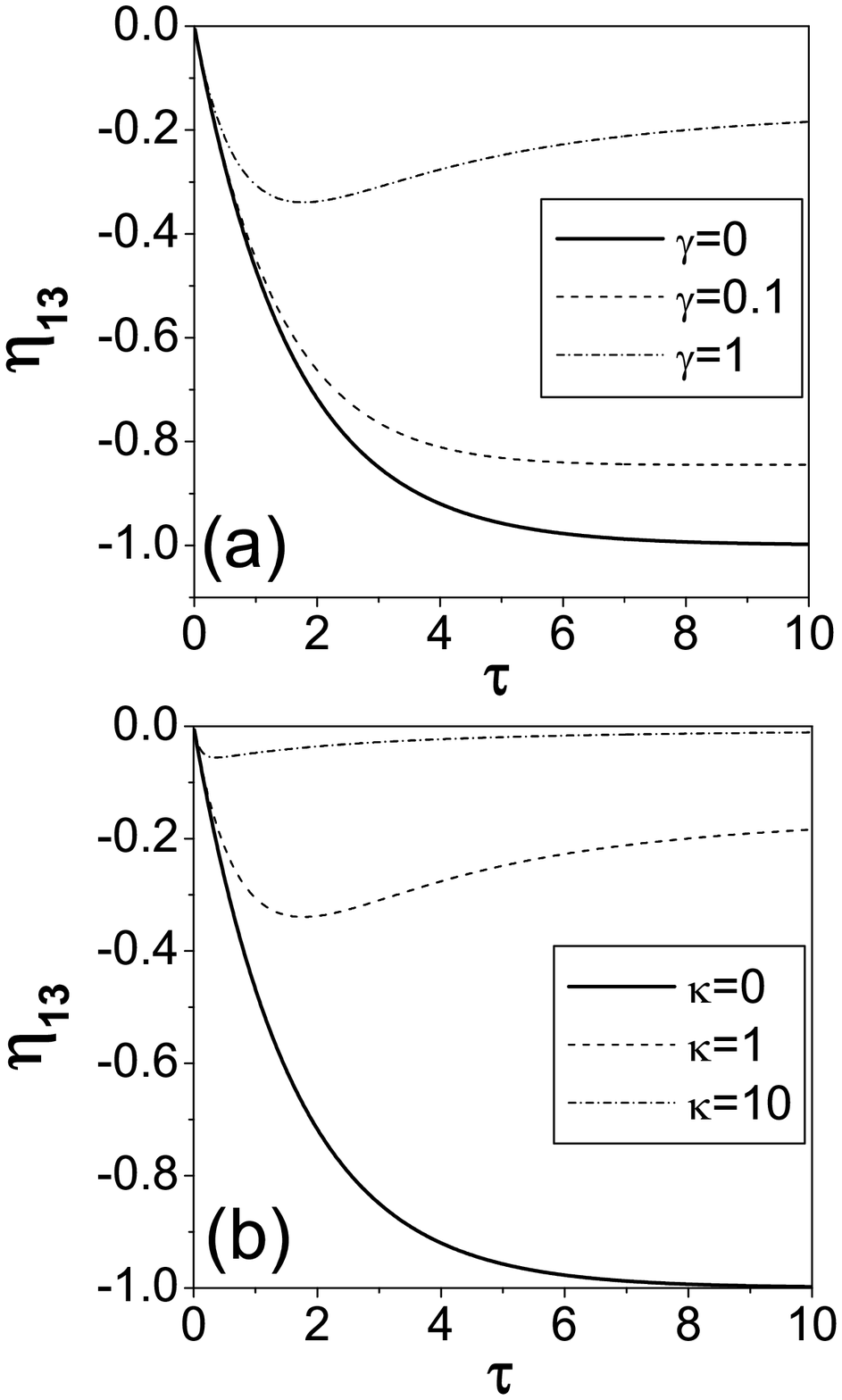}
\caption{Quantum regime  for $\rho=0.2$ and $\delta=5$: minimum
eigenvalue of matrix $S_{13}$ for $\kappa=0$ and different values
of $\gamma$ (a) and for $\gamma=0$ and different values of
$\kappa$ (b).} \label{fig15}
\end{center}
\end{figure}

In Fig. \ref{fig14} and \ref{fig15} are plotted the minimum
eigenvalues of matrices $S_{12}$, $S_{13}$ in the quantum regime, with $\rho=0.2$.
Here the atom-photon entanglement in the state $13$ is robust 
while atom-atom entanglement of the state $12$ is not. The minimum
eigenvalue always remains negative, but it starts from a very small
absolute value and approaches very fast to zero increasing
$\kappa$ and $\gamma$.

\section{Conclusions}\label{s:out}
We have investigated how cavity radiation losses and atomic decoherence
influence the generation of two (atom-atom or atom-radiation) and three mode
entanglement in the collective atomic recoil lasing (CARL) by a Bose-Einstein
condensate driven by a far off-resonant pump laser. The atoms back-scatter
photons from the pump to a weak radiation mode circulating in a ring cavity,
recoiling with opposite momentum $\pm 2\hbar k_p$ along the ring cavity axis.
Our analysis has been focused to the linear regime, in which the ground state
of the condensate remains approximately undepleted and the dynamics is
described by three parametrically coupled boson operators, corresponding to
the radiation mode and two condensates with momentum displaced by $\pm\
2\hbar\vec k_p$.  The problem resembles that of three optical modes generated in
a $\chi^{(2)}$ medium \cite{opa} and thus our results may have a more general
interest also behind the physics of the BEC. We have solved analytically the
dissipative Master equation in terms of the Wigner function and we have
investigated the entanglement properties of the evolved state. We found that
three-mode entanglement as well two-mode atom-atom and atom-photon
entanglement is generally robust against cavity losses and decoherence. The
analysis has been focused of the different dynamical regimes, the high-Q cavity
regime, with low cavity losses, and the superradiant regime in the so-called
'bad-cavity limit'. We have found that entanglement in the high-Q cavity
regime is generally robust against either cavity or decoherence losses. On the
contrary, losses seriously limit atom-atom and atom-radiation number squeezing
production in CARL \cite{Gasenzer}. Concerning the superradiant regime,
atom-atom entanglement in the semi-classical limit is generally more robust
than atom-radiation entanglement in the quantum-limit. Finally, we have proved
that the state generated in the ideal case without losses is fully
inseparable.  We conclude that the present system is a good candidate for the
experimental observation of entanglement in condensate systems since, in
particular, steady-state entanglement may be obtained both between atoms with
opposite momenta  and between atoms and photons.

\section*{Acknowledgments}
This work has been sponsored by INFM and by MIUR. MGAP is research fellow at 
{\em Collegio Alessandro Volta}. We thank A. Ferraro for stimulating discussions.

\appendix

\section{Solution of the Fokker-Planck equation}\label{a:appA}

In order to solve Eq.(\ref{fokker}) for the Green function $G(\mathbf{u},t;\mathbf{u}_{0},0)$
it is helpful to first perform a similarity transformation to
diagonalize the drift matrix $\mathbf{A}$:
\begin{equation}
\tilde{\mathbf{A}}=\mathbf{S}\mathbf{A}\mathbf{S}^{-1}=\mbox{diag}
\{\lambda_{1}\lambda_{2}\lambda_{3}\},\label{diagA}
\end{equation}
where the complex eigenvalues $\lambda_{j}$ of $\mathbf{A}$ (with $j=1,2,3$)
are obtained from the characteristic equation
\begin{equation}
\det(\mathbf{A}-\lambda \mathbf{I})=0,
\label{det}
\end{equation}
$\mathbf{I}$ is the $3\times3$ identity matrix and the columns of
$\mathbf{S}^{-1}$ are the right eigenvectors of $\mathbf{A}$ with
$\det(\mathbf S)=1$.  Solving Eq.(\ref{det}) we obtain
$\lambda_j=i(\omega_j-\delta)-\gamma_+$, where $\omega_j$ are the three roots
of the cubic equation (\ref{cubica}), whereas the eigenvectors of $\mathbf{A}$
corresponding to the $j$-th eigenvalue are
\begin{equation}
\mathbf{a}_{j}^{T}={\cal N}_{j}\left(i\sqrt{\frac{\rho}{2}}
\left(\omega_{j}+\beta\right),
-i\sqrt{\frac{\rho}{2}}\left(\omega_{j}-\beta\right),
-\omega_{j}^{2}+\beta^2 \right),
\end{equation}
where $\beta=1/\rho+i\gamma_-$ and
\begin{equation}
{\cal N}_{1}=\frac{1}{\omega_{2}-\omega_{3}}\;\;\;\;
{\cal N}_{2}=\frac{1}{\omega_{1}-\omega_{3}}\;\;\;\;
{\cal N}_{3}=\frac{1}{\omega_{1}-\omega_{2}}.
\end{equation}
Explicitly calculating the inverse matrix of $S^{-1}$ we have
\begin{equation} \mathbf{S}=\left(\begin{array}{rrr}
i\sqrt{\rho/2}\; (a_{22}a_{23}/{\cal N}_{1})& i\sqrt{\rho/2}\;
(a_{12}a_{13}/{\cal N}_{1})& -{\cal N}_{2}{\cal N}_{3}/{\cal N}_{1}\\
-i\sqrt{\rho/2} \;(a_{23}a_{21}/{\cal N}_{2})&
-i\sqrt{\rho/2} \;(a_{11}a_{13}/{\cal N}_{2})& {\cal N}_{1}{\cal N}_{3}/{\cal N}_{2}\\
i\sqrt{\rho/2} \;(a_{21}a_{22}/{\cal N}_{3})& i\sqrt{\rho/2}\;
(a_{12}a_{11}/{\cal N}_{3})& -{\cal N}_{1}{\cal N}_{2}/{\cal N}_{3},
\end{array}\right)
\end{equation}
where $a_{ij}=(\mathbf{a}_{j})_{i}$.
Now we transform the Fokker-Plank equation (\ref{fokker})
in the new variable $\mathbf{v}\equiv \mathbf{S}\mathbf{u}$. From (\ref{diagA}) we obtain
\begin{eqnarray}
\mathbf{u}'^{T}\mathbf{A}\mathbf{u}&=&\mathbf{u}'^{T}
(\mathbf{S}^{-1}\tilde{\mathbf{A}}\mathbf{S})\mathbf{u}
=\mathbf{v}'^{T}\tilde{\mathbf{A}}\mathbf{v}\label{primo}\\
\mathbf{u}'^{T}\mathbf{D}\mathbf{u}'^{*}&=&
(\mathbf{v}'^{T}\mathbf{S})\mathbf{D}(\mathbf{S}^{T}\mathbf{v}')^*=
\mathbf{v}'^{T}\tilde{\mathbf{D}}\mathbf{v}'^{*},\label{secondo}
\end{eqnarray}
where
$\tilde{\mathbf{D}}\equiv\mathbf{S}\mathbf{D}\mathbf{S}^{\dag}$,
$\mathbf{S}^{\dag}=(\mathbf{S}^{T})^*$ and
$\mathbf{v}'^{T}=\mathbf{u}'^{T}\mathbf{S}^{-1}$.
Using (\ref{primo}) and (\ref{secondo}) Eq. (\ref{fokker}) becomes
\begin{eqnarray}
\frac{\partial \tilde{W}}{\partial \tau}=
-\left(\textbf{v}'^{T}\mathbf{\tilde{A}}\textbf{v} +{\rm c.c.}\right)\tilde{W}
+\textbf{v}'^{T}\tilde{\mathbf{D}}\textbf{v}'^{*}\tilde{W},\label{fokkerv}
\end{eqnarray}
where $\tilde{W}(\mathbf{v},\tau)=W(\mathbf{S}^{-1}\mathbf{v},\tau)$.
Eq.(\ref{fokkerv}) is a linear Fokker-Plank equation with diagonal drift.
Introducing the Fourier transform
\begin{equation}
\tilde{U}(\mathbf{k},\tau)=
\int\frac{d^{2}\mathbf k}{\pi^3}
\tilde W(\mathbf{v})\exp(\mathbf{k}^{*T}\mathbf{v}-\mathbf{k}^T\mathbf{v}^*),
\end{equation}
Eq.(\ref{fokkerv}) becomes
\begin{equation}
\frac{\partial \tilde{U}}{\partial \tau}=\left(
\mathbf{k}^{*T}\mathbf{\tilde{A}}\mathbf{k}'^* + \mathbf{k}^{T}\mathbf{\tilde{A}}^*\mathbf{k}'
\right) \tilde{U}
-\left(\mathbf{k}^{*T}\tilde{\mathbf{D}}\mathbf{k}\right)\tilde{U}.
\label{fokkerft}
\end{equation}
where
\begin{equation}
\textbf{k}^T= \left( k_{1}, k_{2}, k_{3}\right)
\qquad \mathbf{k}'^{T}=\left(
\frac{\partial}{\partial k_{1}},
\frac{\partial}{\partial k_{2}},
\frac{\partial}{\partial k_{3}}\right).
\end{equation}
The Fourier transform of initial condition of the Green function
$\tilde{G}(\mathbf{v},0;\mathbf{S}\mathbf{u}_{0},0)=\delta^{3}(\mathbf{v}-\mathbf{S}\mathbf{u}_{0})$
is 
\begin{equation}
\tilde{U}(\mathbf{k},0)=
\exp\left[\mathbf{k}^{*T}\mathbf{S}\mathbf{u}_{0}
-\mathbf{k}^{T}(\mathbf{S}\mathbf{u}_{0})^{*}\right].
\label{incondition}
\end{equation}
Eq.(\ref{fokkerft}) is now solved using the method of the characteristics.
Since $\tilde{\mathbf{A}}$ is diagonal the subsidiary equations are
\begin{equation}
\frac{d\tau}{1}=
\frac{d k^*_1}{-\lambda_1 k^*_2}=
\frac{d k^*_2}{-\lambda_2 k^*_2}=
\frac{d k^*_3}{-\lambda_3 k^*_3}=
\frac{d\tilde{U}}{(-\mathbf {k}^{*T}\tilde{\mathbf{D}}\mathbf{k})\tilde{U}}
\end{equation}
and have solutions
\begin{equation}
\mathbf{k}=e^{-\tilde{\mathbf{A}}^{*}\tau}\mathbf{c}={\rm constant}\label{constant}.
\end{equation}
Then
\begin{equation}
\frac{d\tilde{U}}{\tilde{U}}=-\mathbf{k}^{*T}\tilde{\mathbf{D}}\mathbf{k}d\tau=-\mathbf{c}^{*T}
\left(
e^{-\tilde{\mathbf{A}}\tau}\tilde{\mathbf{D}}
e^{-\tilde{\mathbf{A}}^*\tau}\right)\mathbf{c}d\tau=
-\mathbf{c}^{*T}\left[\tilde{\mathbf{D}}_{ij}
e^{-(\lambda_{i}+\lambda_{j}^*)\tau}\right]\mathbf{c}d\tau,
\end{equation}
where $(\mathbf{B}_{ij})$ denotes the matrix with elements $\mathbf{B}_{ij}$, and we find,
using Eq. (\ref{constant}),
\begin{equation}
\ln\tilde{U}=
\mathbf{k}^{*T}\left\{\frac{\tilde{\mathbf{D}}_{ij}}
{\lambda_{i}+\lambda_{j}^{*}}\left[1-
e^{(\lambda_{i}+\lambda_{j}^{*})\tau}\right]\right\}\mathbf{k}+\mbox{constant}\:.
\end{equation}
It follows that
\begin{equation}
\tilde{U}\exp\left\{\mathbf{k}^{*T}\tilde{\mathbf{Q}}\mathbf{k}\right\}
=\mbox{constant},\label{Uconstant}
\end{equation}
where $\tilde{\mathbf{Q}}$ is the $3\times3$ matrix with elements
\begin{equation}
\tilde{\mathbf{Q}}_{ij} \equiv -\frac{\tilde{D}_{ij}}
{\lambda_{i}+\lambda_{j}^{*}}\left[1-
e^{(\lambda_{i}+\lambda_{j}^{*})\tau}\right]=
\int_0^\tau d\tau'\tilde{D}_{ij}e^{(\lambda_i+\lambda_j^*)\tau'}.
\end{equation}
Thus, from Eqs. (\ref{constant}) and (\ref{Uconstant}), the solution
for $\tilde{U}$ takes the general form
\begin{equation}
\tilde{U}(\mathbf{k},\tau)=\Phi(e^{\tilde{\mathbf{A}}^*\tau}\mathbf{k})
\exp\{-\mathbf{k}^{*T}\tilde{\mathbf{Q}}\mathbf{k}\}
\end{equation}
where $\Phi$ is an arbitrary function. Choosing $\Phi$ to match
the initial condition (\ref{incondition}), we find
\begin{equation}
\tilde{U}(\mathbf {k},\tau)=
\exp\left\{\mathbf {k}^{*T}
(\mathbf{S}e^{\mathbf{A}\tau}\mathbf{u}_{0})
-\mathbf{k}^{T}(\mathbf{S}e^{\mathbf{A}\tau}\mathbf{u}_{0})^*\right\}
\exp\left\{-\mathbf{k}^{*T}\tilde{\mathbf{Q}}\mathbf{k}\right\}.
\end{equation}
In the argument of the first exponential on the right-hand side we
have used (\ref{diagA}) to write
$\exp(\tilde{\mathbf{A}}\tau)\mathbf{S}=\mathbf{S}\exp(\mathbf{A}\tau)$.
Inverting the Fourier transform we obtain
\begin{equation}
\tilde{G}(\mathbf{v},\tau;\mathbf{S}\mathbf{u}_{0},0)
=\frac{1}{\pi^{3}\det\mathbf{\tilde{Q}}}
\exp\left\{\left(\mathbf{v}-\mathbf{S}e^{\mathbf{A}\tau}\mathbf{u}_{0}\right)^\dag
\mathbf{\tilde{Q}}^{-1}\left(\mathbf{v}-\mathbf {S}e^{\tilde{\mathbf{A}}\tau}
\mathbf{u}_{0}\right)\right\}
\end{equation}
and so transforming back the variables
\begin{equation}
G(\mathbf{u},\tau;\mathbf{u}_{0},0)=\frac{1}{\pi^{3}\det\mathbf{Q}}
\exp\left\{\left(\mathbf{u}-e^{\mathbf{A}\tau}\mathbf{u}_{0}\right)^\dag
\mathbf{Q}^{-1}\left(\mathbf{u}-e^{\mathbf{A}\tau}\mathbf{u}_{0}\right)\right\}.
\end{equation}
where
\begin{equation}
\mathbf{Q}=\mathbf{S}^{-1}\hat\mathbf{Q}\left(\mathbf{S}^{-1}\right)^\dag
=\int_0^\tau d\tau' e^{\mathbf{A}\tau'}\mathbf{D}\left(e^{\mathbf {A}\tau'}\right)^\dag.
\label{Q2}
\end{equation}

\section{Elements of the matrices ${\mathbf M }$, Eq. (\ref{eAt}), 
and ${\mathbf C}$, Eq. (\ref{cov})}\label{a:appB}
The expressions of the functions $f_{ij}$ which appear as 
elements of the matrix $\mathbf{M}$, Eq. (\ref{eAt}),
are
\begin{eqnarray}
f_{11}(\tau)&=&e^{-(\gamma_++i\delta)\tau}\sum_{j=1}^3[(\omega_j-\alpha)(\omega_j+\beta)-\rho/2]
\frac{e^{i\omega_j\tau}}{\Delta_j}\label{f11}\\
f_{22}(\tau)&=&e^{-(\gamma_++i\delta)\tau}\sum_{j=1}^3[(\omega_j-\alpha)(\omega_j-\beta)+\rho/2]
\frac{e^{i\omega_j\tau}}{\Delta_j}\label{f22}\\
f_{33}(\tau)&=&e^{-(\gamma_++i\delta)\tau}\sum_{j=1}^3(\omega_j^2-\beta^2)
\frac{e^{i\omega_j\tau}}{\Delta_j}\label{f33}\\
f_{12}(\tau)&=&-\frac{\rho}{2}e^{-(\gamma_++i\delta)\tau}\sum_{j=1}^3
\frac{e^{i\omega_j\tau}}{\Delta_j}\label{f12}\\
f_{13}(\tau)&=&-i\sqrt{\frac{\rho}{2}}e^{-(\gamma_++i\delta)\tau}\sum_{j=1}^3(\omega_j+\beta)
\frac{e^{i\omega_j\tau}}{\Delta_j}\label{f13}\\
f_{23}(\tau)&=&i\sqrt{\frac{\rho}{2}}e^{-(\gamma_++i\delta)\tau}\sum_{j=1}^3(\omega_j-\beta)
\frac{e^{i\omega_j\tau}}{\Delta_j}\label{f23}
\end{eqnarray}
where $\alpha=\delta+i(\kappa-\gamma_+)$, $\beta=1/\rho+i\gamma_-$,
$\Delta_j=(\omega_j-\omega_k)(\omega_j-\omega_m)$ (with $j\neq k\neq m$) and $\omega_1$,
$\omega_2$ and $\omega_3$ are the roots of the cubic Eq.(\ref{cubica}). It is possible to show that
$f_{ij}(0)=\delta_{ij}$ in order to satisfy the initial condition $\mathbf{M}(0)=\mathbf{I}$.

The explicit components of the covariance matrix
\begin{equation}
\mathbf{C}(\tau)=\mathbf{Q}(\tau)+\frac{1}{2}\mathbf{M}(\tau)\mathbf{M}^{\dag}(\tau),\nonumber
\end{equation}
where $\mathbf{M}$ and $\mathbf{Q}$ are defined in (\ref{eAt}) and (\ref{Q}), are
\begin{eqnarray}
C_{11}(\tau)&=&\int_0^\tau d\tau'\left\{
\gamma_1|f_{11}|^2+\gamma_2|f_{12}|^2+\kappa|f_{13}|^2\right\}
+\frac{1}{2}\left(|f_{11}|^2+|f_{12}|^2+|f_{13}|^2\right)\label{C11}\\
C_{22}(\tau)&=&\int_0^\tau d\tau'\left\{
\gamma_1|f_{12}|^2+\gamma_2|f_{22}|^2+\kappa|f_{23}|^2\right\}
+\frac{1}{2}\left(|f_{12}|^2+|f_{22}|^2+|f_{23}|^2\right)\label{C22}\\
C_{33}(\tau)&=&\int_0^\tau d\tau'\left\{
\gamma_1|f_{13}|^2+\gamma_2|f_{23}|^2+\kappa|f_{33}|^2\right\}
+\frac{1}{2}\left(|f_{13}|^2+|f_{23}|^2+|f_{33}|^2\right)\label{C33}\\
C_{12}(\tau)&=&\int_0^\tau d\tau'\left\{
-\gamma_1 f_{11}f_{12}^* +\gamma_2f_{12}f_{22}^*+\kappa f_{13}f_{23}^*\right\}
+\frac{1}{2}\left(-f_{11}f_{12}^* +f_{12}f_{22}^*+f_{13}f_{23}^*\right)\label{C12}\\
C_{13}(\tau)&=&\int_0^\tau d\tau'\left\{
 \gamma_1 f_{11}f_{13}^* -\gamma_2f_{12}f_{23}^*+\kappa f_{13}f_{33}^*\right\}
+\frac{1}{2}\left(f_{11}f_{13}^* -f_{12}f_{23}^*+f_{13}f_{33}^*\right)\label{C13}\\
C_{23}(\tau)&=&\int_0^\tau d\tau'\left\{
-\gamma_1 f_{12}f_{13}^* -\gamma_2f_{22}f_{23}^*+\kappa f_{23}f_{33}^*\right\}
+\frac{1}{2}\left(-f_{12}f_{13}^* -f_{22}f_{23}^*+f_{23}f_{33}^*\right)\label{C23}
\end{eqnarray}
with $C_{ij}=C_{ji}^*$.

In the special case $\gamma_+=\gamma=\kappa$ and $\gamma_-=0$,
$f_{ij}=e^{-\gamma\tau}f_{ij}^{(0)}$, where $f_{ij}^{(0)}$ is the solution
without losses. As shown in Ref.\cite{PRA}, they satisfy the following
relations:
\begin{eqnarray}
|f_{13}^{(0)}|^2+1&=&|f_{23}^{(0)}|^2+|f_{33}^{(0)}|^2  \label{r1}\\
|f_{11}^{(0)}|^2-1&=&|f_{12}^{(0)}|^2+|f_{13}^{(0)}|^2  \label{r2}\\
|f_{12}^{(0)}|^2+1&=&|f_{22}^{(0)}|^2+|f_{23}^{(0)}|^2  \label{r3}\\
f_{11}^{(0)}(f_{13}^{(0)})^*&=&-f_{12}^{(0)}(f_{23}^{(0)})^*+f_{13}^{(0)}(f_{33}^{(0)})^*
\label{r4}\\
-f_{11}^{(0)}(f_{12}^{(0)})^*&=&f_{12}^{(0)}(f_{22}^{(0)})^*+f_{13}^{(0)}(f_{23}^{(0)})^*
\label{r5}\\
-f_{12}^{(0)}(f_{13}^{(0)})^*&=&-f_{22}^{(0)}(f_{23}^{(0)})^*+f_{23}^{(0)}(f_{33}^{(0)})^*
\label{r6}
\end{eqnarray}
Using Eqs.(\ref{r1})-(\ref{r3}) in Eq.(\ref{C11})-(\ref{C33}) and  $C_{ii}=1/2+
\langle{\hat n_i}\rangle$, we obtain
that
\begin{equation}
\langle{\hat n_1}\rangle=\langle{\hat n_2}\rangle+\langle{\hat n_3}\rangle
\label{const}
\end{equation}
and
\begin{equation}
\frac{d\langle{\hat n_i}\rangle}{d\tau}=\frac{d\langle{\hat n_i^{(0)}}\rangle}{d\tau}e^{-2\gamma\tau},
\label{damped}
\end{equation}
where $\langle{\hat n_i^{(0)}}\rangle$ are the expectation values of the occupation numbers of the three modes
in the ideal case without losses.

\par

\end{document}